\documentclass[prx,two column,show pacs,show keys,superscript address,superscript reference,floatfix]{revtex4-1}
\usepackage[colorlinks=true,citecolor=blue,linkcolor=blue,urlcolor=red]{hyperref}

\newcommand{\ket}[1]{\ensuremath{\left|{#1}\right\rangle}}

\usepackage[sort&compress]{natbib}
\usepackage{graphicx,times}
\usepackage{color}
\usepackage{sansmath}
\usepackage{dsfont}
\usepackage{amssymb}
\usepackage{amsthm}

\usepackage{subeqnarray}
\usepackage{bm}
\usepackage{diagbox}
\usepackage{makecell}
\usepackage{amsmath}
\newcommand{\mathbbm}[1]{\text{\usefont{U}{bbm}{m}{n}#1}}
\newtheorem*{definition}{Definition}

\begin{document}

\title{Digital-Analog Quantum Computation}

\author{Adrian Parra-Rodriguez}
\affiliation{Department of Physical Chemistry, University of the Basque Country UPV/EHU, Apartado 644, E-48080 Bilbao, Spain}

\author{Pavel Lougovski}
\affiliation{Quantum Information Science Group, Oak Ridge National Laboratory, Oak Ridge, Tennessee 37831, USA}

\author{Lucas Lamata}
\affiliation{Department of Physical Chemistry, University of the Basque Country UPV/EHU, Apartado 644, E-48080 Bilbao, Spain}
\affiliation{Departamento de F\'isica At\'omica, Molecular y Nuclear, Universidad de Sevilla, 41080 Sevilla, Spain}

\author{Enrique Solano}
\affiliation{Department of Physical Chemistry, University of the Basque Country UPV/EHU, Apartado 644, E-48080 Bilbao, Spain}
\affiliation{IKERBASQUE, Basque Foundation for Science, Maria Diaz de Haro 3, 48013 Bilbao, Spain}
\affiliation{International Center of Quantum Artificial Intelligence for Science and Technology~(QuArtist)\\
	and Physics Department, Shanghai University, 200444 Shanghai, China}

\author{Mikel Sanz}
\email{mikel.sanz@ehu.es}
\affiliation{Department of Physical Chemistry, University of the Basque Country UPV/EHU, Apartado 644, E-48080 Bilbao, Spain}

\begin{abstract}
Digital quantum computing paradigm offers highly-desirable features such as universality, scalability, and  quantum error correction. However, physical resource requirements to implement useful error-corrected quantum algorithms are prohibitive in the current era of NISQ devices. As an alternative path to performing universal quantum computation, within the NISQ era limitations, we propose to merge digital single-qubit operations with analog multi-qubit entangling blocks in an approach we call digital-analog quantum computing (DAQC). Along these lines, although the techniques may be extended to any resource, we propose to use unitaries generated by the ubiquitous Ising Hamiltonian for the analog entangling block and we prove its universal character. We construct explicit DAQC protocols for efficient simulations of arbitrary inhomogeneous Ising, two-body, and $M$-body spin Hamiltonian dynamics by means of single-qubit gates and a fixed homogeneous Ising Hamiltonian. Additionally, we compare a sequential approach where the interactions are switched on and off (stepwise~DAQC) with an always-on multi-qubit interaction interspersed by fast single-qubit pulses (banged DAQC). Finally, we perform numerical tests comparing purely digital schemes with DAQC protocols, showing a remarkably better performance of the latter. The proposed DAQC approach combines the robustness of analog quantum computing with the flexibility of digital methods.
\end{abstract}

\maketitle
\section{Introduction}
Quantum information science has flourished in recent years as a new paradigm promising to outperform certain classical tasks such as computation, simulations, and communications among others. More specifically, quantum computers (QC)~\cite{Benioff_1980,Benioff_1982,Feynman_1982} are believed to be faster than their classical counterparts in factoring prime numbers~\cite{Shor_1996}, or searching databases~\cite{Grover_1996}. From a theoretical computer science point of view, a universal quantum computer can run any algorithm processable by a quantum Turing machine~\cite{Deutsch_1985}, in other words, it can implement an arbitrary unitary evolution. In physical terms, an ideal universal quantum computer can implement an arbitrary Hamiltonian acting on an infinite Hilbert space. However, a realistic quantum computer comprises a finite number of resources, and hence can only perform unitary operations within certain constraints.

The complexity of classical simulations of many-body quantum systems typically grows exponentially with the dimension of the system. This was first recognized by Richard Feynman in a seminal paper from 1982~\cite{Feynman_1982}, in which he proposed as an efficient solution the simulation of these problems employing another fully-controllable quantum system with a similar encoded dynamics~\cite{Johnson_QS_2014}. This was the origin of what is now called {\it analog quantum computing} (AQC). Later, with the emergence of the first quantum algorithms offering speedups over their classical counterparts~\cite{Shor_1996,Grover_1996}, quantum computing became the most promising application of quantum mechanics and information. The discovery of universal sets of quantum gates and quantum error correction~\cite{Shor_1995_QEC,Bennett_1996_QEC,Kitaev_2003_FTQC} provided a clear roadmap towards a scalable QC mimicking the history of classical computers. This approach is called {\it digital quantum computation} (DQC)~\cite{Deutsch_1989,Yao_1993}, based on an algorithmic sequence of one-qubit and two-qubit gates~\cite{Deutsch_1985}. However, a practical universal digital quantum computer is considered to be so resource-consuming that the implementation of useful applications may be shifted decades into the future. It has been suggested that quantum control techniques \cite{DAlessandro_2007,Dong_2010} can be used to improve the fidelity of the quantum gates. However, they require classical optimization algorithms, which are by themselves hard problems to solve \cite{Childs_2018,Song_2018}. Furthermore, they do not consider all possible errors in the system \cite{Ramakrishna_1996} yet. In this context, the ability of quantum systems to solve problems beyond the reach of any current classical computer is known as quantum supremacy~\cite{Preskill_2012_QSup,Preskill_2018}. This has been proposed for various artisanal problems, such as boson sampling \cite{Aaronson_2013_BSamp,Gard_2013_BosonSamp} and quantum speckle~\cite{Neill_2018_QSpec,Boixo_2018_QSup_QSpec}, but still not achieved. 

The simplest approach to perform quantum simulations is the use of a controllable quantum system whose effective dynamics is similar to the one of the desired model. Such single-purpose devices are called {\it analog quantum simulators} (AQS). In this sense, there are many proposals ranging from the quantum Rabi model \cite{Ballester_2012,JB_2017,Pedernales_2015_Rabi,Lv_2018_Rabi} and Casimir physics \cite{Felicetti_2014,Rossatto_2016,Sanz_2018}, to Jaynes-Cummings and Rabi lattices \cite{CarusottoCiuti_2013_Review,Hartmann_2006,Hartmann_2007,Greentree_2006}. In 1996~\cite{Lloyd_1996}, Lloyd proved with the help of the Suzuki--Trotter decomposition \cite{Suzuki_1976} that {\it digital quantum simulators} (DQS), whose evolutions are decomposed in a universal set of quantum gates acting on a register of qubits, can simulate efficiently any quantum system. Innovative experiments were performed in those lines in superconducting circuits \cite{Barends_2015,Salathe_2015,Barends_2016,Klco_2018} and ion traps \cite{Lanyon_2011_DQS,Martinez_2016}. An alternative paradigm to do quantum simulations, called {\it digital-analog quantum simulations} (DAQS), makes use of both digital and analog blocks in order to exploit their intrinsic versatility and complexity~\cite{Mezzacapo_2014_DAQS, GarciaAlvarez_2015, Mezzacapo_2015, Arrazola_2016, Sweke_2016, GarciaAlvarez_2016, Langford_2017, Lamata_2018_DAQS}.

Here, we propose the concept of {\it digital-analog quantum computation}~(DAQC) as a method to reduce the number of gates needed to perform quantum algorithms, in the current spirit of near-term intermediate-scale quantum computation~(NISQ). The DAQC paradigm requires the implementation of entangling multipartite evolutions and fast single-qubit gates. Earlier efforts have already proven the universality of such schemes~\cite{Dodd_2002_UniversalQCandS,Masanes_2002,Bennett_2002,Jane_2003}, extending the ideas triggered by Lloyd \cite{Lloyd_1996}. In our DAQC approach, we propose a sequence of entangling time slices, called analog blocks, and fast single-qubit rotations (SQRs), which belong to our class of digital steps. Moreover, we develop constructive protocols for simulating arbitrary inhomogeneous two-body and $M$-body spin Hamiltonians. An important source of errors when performing quantum algorithms appears by turning on and off interacting Hamiltonians. In order to mitigate these errors, we introduce the concept of banged digital-analog quantum computing (bDAQC) to improve the stepwise DAQC (sDAQC). In this manner, the proposed algorithms do not require to halt and activate the analog blocks, while the only required pulses are single-qubit gates. Furthermore, we perform numerical studies of ideal and realistic multi-qubit models supporting our theoretical results.

\section{Analog, Digital, and Digital-Analog Quantum Computing}

The concepts of digital and analog quantum computing are broadly used in an informal manner but, to the best of our knowledge, they lack of a broadly-accepted formal definition, allowing for a classification of algorithms.  As the main aim of this manuscript is to describe the DAQC paradigm, proper definitions are of crucial importance. Let us start by introducing the concepts of quantum gates, digital and analog blocks on $N$-qubit systems. 
\begin{definition}[Quantum gate]
A quantum gate is a fixed unitary evolution $U_n$,  $U_n\in\mathcal{B}((\mathbbm{C}^2)^{\otimes n})$. 
\end{definition}
\begin{definition}[Digital block]
A $k$-parametric continuous family of unitary operators $U_n(\vec{\phi})$, with $\phi_l\in \mathcal{I}_l(\mathbbm{R})$ and $1\leq l\leq k$,  comprises a {\it digital block} if it is equivalent to a fixed unitary evolution $U_n$ up to a set of local rotations $W_i(\vec{\phi})$, i.e.,  $U_n (\vec{\phi}) = \left (\bigotimes_{i}^n W_{i}(\vec{\phi})\right ) U_n$.
\end{definition}
See, that both parameter-fixed entangling quantum gates and single qubit rotations with arbitrary angle are digital blocks.
\begin{definition}[Analog block]
We call {\it analog block} a $k$-parameter-dependent entangling unitary evolutions $V_n(\vec{\phi})$ with a semigroup structure $V_n(\vec{\phi})=V_n(\vec{\phi}_1)V_n(\vec{\phi}_2)$; $\vec{\phi}=\vec{\phi}_1+\vec{\phi}_2$. For $k=0$, it obviously becomes a quantum gate.
\end{definition}
Under these definitions, for instance, $U_n=e^{i\frac{\pi}{4}\sigma_z^1 \sigma_z^2}$ is a quantum gate, both $U_n(\phi)=(e^{i\phi_1 \sigma_z^1}\otimes e^{i\phi_2 \sigma_z^2})e^{i\frac{\pi}{4}\sigma_z^i \sigma_z^j}$ and $W_i(\phi)=e^{i\phi \sigma_z^1}$ are digital blocks, and $V_n(\phi)=e^{i\phi \sigma_z^i \sigma_z^j}$ is an analog block.

Let us remark that, differently to the analog block, the entanglement generated by the digital block is the same for all values of the parameters.  We call a quantum protocol a {\it digital quantum algorithm}, when it makes use only of digital blocks (usually with a small number of quantum gates), whereas an {\it analog quantum algorithm} consists on a single analog block for different values of the parameters. Naturally, a digital-analog protocol contains both digital and analog blocks. In this paper, we will constrain the digital blocks to arbitrary single qubit rotations, such that our total evolution can be written as $\prod_{j} \left (U_j(\vec{\phi}_j)\left [\bigotimes_i W_{i}^{(j)}(\vec{\alpha}_{ji})\right ]\right)$. 

For the sake of clarity, Reference \cite{Lamata_2018_DAQS} and references thereof which use the terminology digital-analog quantum simulations are, based on the aforementioned definitions, purely digital protocols employing multi-qubit fixed-phase gates, e.g.,  M\"olmer-S\"orensen gates.

\section{Digital-Analog Quantum Computation}
\label{Sec:DAQC}
As it was elegantly proven in Ref. \cite{Dodd_2002_UniversalQCandS}, almost any two-body Hamiltonian is universal. In this Article, we will focus on two paradigmatic models, ubiquitous in quantum platforms, to exemplify the DAQC paradigm, but it could be straightforwardly extended to other specific situations.  We will employ as analog blocks either a homogeneous nearest-neighbor or a homogeneous all-to-all two-body Ising Hamiltonian. Using one of these evolutions, together with single-qubit rotations, we constructively prove their universality, i.e.,  any unitary can be arbitrarily close simulated employing these resources. We show the protocols, sometimes optimal, to generate increasingly complex families of Hamiltonians with relevance in several fields. The roadmap towards this complexity comprises the protocols to generate: (i) arbitrary inhomogeneous two-body Ising Hamiltonians, (ii) an inhomogeneous two-body Hamiltonian, for which the non-commutation of the terms requires to superimpose Trotterization to the algorithm, and (iii) an arbitrary $M$-body Hamiltonian, with polynomial number of resources in the number of spins for fixed $M$. Finally, we perform several numerical simulations to check the advantages in fidelity and time of DAQC with respect to DQC.

Our DAQC algorithms consider three basic ingredients that were also required in  Ref. \cite{Dodd_2002_UniversalQCandS}. Firstly, we need to exactly evolve for a time $t$ with a Hamiltonian $H$, such that combined with local rotations we have an evolution with $H'=RHR^\dagger$. We will later consider the effect of limited precision in the physical implementation of engineered times. Secondly, we must do a Trotter decomposition $e^{-it H}=(e^{-iH_1 t /n_T}e^{-iH_2 t /n_T})^{n_T}+O(t^2/n_T)$ in order to simulate Hamiltonians $H=H_1+H_2$ with non-commuting terms, i.e.,  $[H_1,H_2]\neq0$. The error introduced through this approximation is of second order in the time error $\Delta t$. We consider further improved decompositions like the symmetrized Trotter decomposition, which reduces the error to $\mathcal{O}(\Delta t)^3$. Finally, given that we can evolve with Hamiltonian $H$ we can control the time parameter to evolve with $\lambda H$, where $\lambda>0$ is a continuous positive parameter.

\subsection{Ising Model}
\label{Subsec:Ising_Model}
Let us start illustrating the problem with the implementation of the inhomogeneous all-to-all (ATA) two-body Ising model described with Hamiltonian $H_{ZZ}=\sum_{j<k}^{N}g_{ij}\sigma_z^{j}\sigma_z^{k}$. We use as the elementary analog block the unitary evolution $U_{zz}(t)=e^{i H_{zz} t}$ of the homogeneous ATA two-body Ising model whose Hamiltonian reads
\begin{equation}\label{eq:Ham_sz_sz_base}
H_{zz}=g\sum_{j<k}^{N}\sigma_z^{j}\sigma_z^{k},
\end{equation}
with independent time parameter $t$ and a fixed coupling strength $g$. We set $\hbar=1$ in the whole article. The only digital blocks employed are single qubit rotations around the $x$-axis with the continuous range of phases $\theta\in[0,2\pi]$. The protocol can be straightforwardly modified to use as an analog block a general fixed inhomogeneous evolution with Hamiltonian $\bar{H}_{ZZ}=\sum_{j<k}^{N}\bar{g}_{jk}\sigma_z^{j}\sigma_z^{k}$. Such Hamiltonians naturally appear in various quantum platforms, e.g.,  the coupling parameters scale typically as $\bar{g}_{jk}\propto 1/|j-k|^{\alpha}$, with $0<\alpha<3$ in ion-trap setups \cite{Porras_2004}. In effective Hamiltonian models, in which qubits are coupled through linear multi-mode systems \cite{LehmbergI_1970,LehmbergII_1970}, more complex coupling distributions naturally emerge and could also be tuned or designed \cite{GonzalezTudela_2015,Solgun_2017}. A different decomposition into most entangling Hamiltonians and single qubit gates has been proven to be a universal quantum machine by Dodd {\it et al.} in Ref. \cite{Dodd_2002_UniversalQCandS}. The Ising Hamiltonian is an example of a universal Hamiltonian under the digital-analog paradigm, since it allows us to construct a universal ZZ gate \cite{vandenNest_2008} between any two arbitrary qubits, see Appendix \ref{Appsec:Universality}.
\begin{figure}[h]
	\includegraphics[width=.9\linewidth]{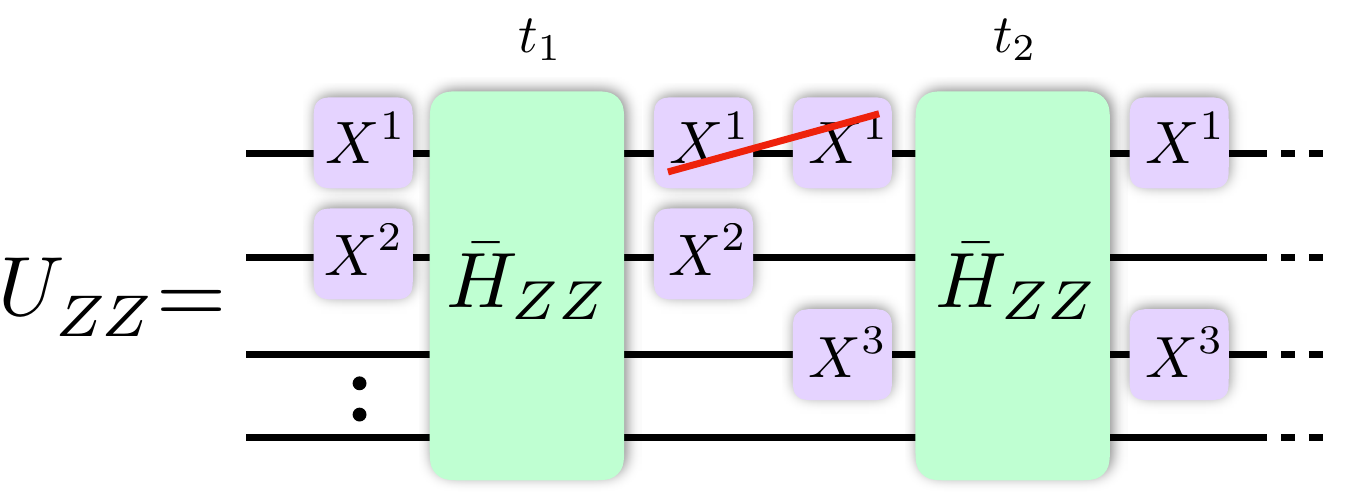}
	\caption{\label{fig:HzzI_bHzzIsDAQC} \textbf{Algorithm to simulate the general inhomogeneous Ising model from a fixed one}. Each time step evolution $t_\alpha$ is sandwiched by a pair of single qubit gates ($X^i\equiv\sigma_x^i$) applied to qubits $(n,m)$, with $\alpha(n,m)$. Optimal sequences of SQRs can be used to simplify the number of pulses.}
\end{figure}
The target Hamiltonian evolves according to the unitary $U_{ZZ}=e^{i t_F H_{ZZ}}$, where $H_{ZZ}=\sum_{j<k}g_{jk}\sigma_z^j\sigma_z^k$ and $t_F$ the final time. The task consists on finding a mapping between $g_{jk}t_F$ and $g t_{nm}$ by slicing the homogeneous time evolution into (at most) $N(N-1)/2$ analog blocks of different time lengths $t_{nm}$, sandwiched by local rotations $\sigma_x^n\sigma_x^m$,
\begin{eqnarray}
H_{ZZ}&=& \sum_{j<k}^N g_{jk}\sigma_z^j\sigma_z^k=\frac{g}{t_F}\sum_{j<k}^N \sum_{n<m}^N t_{nm}\sigma_x^n \sigma_x^m \sigma_z^j \sigma_z^k \sigma_x^n \sigma_x^m\label{eq:H_I_ZZ}\nonumber\\
&=& \frac{g}{t_F} \sum_{j<k}^N \sum_{n<m}^N t_{nm}(-1)^{\delta_{nj}+\delta_{nk}+\delta_{mj}+\delta_{mk}}\sigma_z^j \sigma_z^k,
\end{eqnarray}
as depicted in Fig. \ref{fig:HzzI_bHzzIsDAQC}. Engineering the Hamiltonian reduces the problem of finding the time durations $t_{nm}$ of the analog block evolutions to a matrix-inversion problem. By vectorizing the pairs of indices $(n,m)\rightarrow\alpha=N(n-1)-n(n+1)/2+m$ and $(j,k)\rightarrow\beta=N(j-1)-j(j+1)/2+k$ and write the signs matrix 
\begin{equation}
\mathsf{M}_{\alpha\beta}=(-1)^{\delta_{nj}+\delta_{nk}+\delta_{mj}+\delta_{mk}},\label{eq:Ising_Malphabeta}
\end{equation}
where the inverted relations are $j=1+\left[\frac{\beta}{N}\right]$ and 
$k=\beta+\frac{1}{2}\left(1+\left[\frac{\beta}{N}\right]\right)\left(2+\left[\frac{\beta}{N}\right]\right)-N\left[\frac{\beta}{N}\right]$ and $n=1+\left[\frac{\alpha}{N}\right]$, $m=\alpha+\frac{1}{2}\left(1+\left[\frac{\alpha}{N}\right]\right)\left(2+\left[\frac{\alpha}{N}\right]\right)-N\left[\frac{\alpha}{N}\right]$.

The matrix $\mathsf{M}_{\alpha\beta}$ has three degenerate eigenvalues, namely, $\lambda_1=N(N-9)/2+8$, $\lambda_2=2(4-N)$ and $\lambda_3=4$ with degeneracies $1$, $N-1$, and $N(N-1)/2-N$, respectively. Consequently, it is a non-singular matrix $\forall N\in \mathbb{Z}-\{4\}$. The corner case $N=4$ requires the use of a slightly modified set of SQRs, e.g.,  single $\sigma_x$-rotations per site is sufficient for a NN Hamiltonian. The total unitary evolution is $U_{ZZ}(t_F) = e^{i t_F \sum_{\beta}H_{ZZ}^\beta }$ where 
\begin{equation}
H_{ZZ}^\beta = g_\beta \sigma_z^{j(\beta)}\sigma_z^{k(\beta)},\label{eq:H_beta}
\end{equation}
and $g_\beta= t_\alpha \mathsf{M}_{\alpha\beta}(g/t_F)$. 
\begin{table}[h]
	\begin{tabular}{l|c|c|c|}
		\diagbox{$\alpha\rightarrow(n,m)$}{$\beta\rightarrow (j,k)$}  & 1$\rightarrow \sigma_z^{1}\sigma_z^{2}$ & 2$\rightarrow \sigma_z^{1}\sigma_z^{3}$ & 3$\rightarrow \sigma_z^{2}\sigma_z^{3}$ \\\hline
		\text{SQRs}\,1$\rightarrow \sigma_x^{1}\sigma_x^{2}$ & $t_1$  & $-t_1$ & $-t_1$\\\hline
		\text{SQRs}\,2$\rightarrow \sigma_x^{1}\sigma_x^{3}$ & $-t_2$  & $t_2$ & $-t_2$\\\hline
		\text{SQRs}\,3$\rightarrow \sigma_x^{2}\sigma_x^{3}$ & $-t_3$  & $-t_3$ & $t_3$\\\hline
	\end{tabular}
	\caption{\label{tab:Ising_M_ab_t_a}{\bf Matrix elements of $t_\alpha \mathsf{M}_{\alpha\beta}$ coupling sign constants for the $N=3$ qubits case.} By performing three pairs of single qubit rotations sandwiching the interaction base Hamiltonian (\ref{eq:Ham_sz_sz_base}), we derive independent equations for all coupling terms $g_{jk}$ in the simulated Hamiltonian (\ref{eq:H_I_ZZ}).}
\end{table}
An example of elements $t_\alpha \mathsf{M}_{\alpha\beta}$ is shown in Table \ref{tab:Ising_M_ab_t_a} for a three-qubit case. Solving the linear problem, we find the times  $t_\alpha= \mathsf{M}_{\alpha\beta}^{-1}g_\beta(t_F/g)$ required for each analog block to evolve interleaved by the pairs of single qubit rotations, as shown in Fig. \ref{fig:HzzI_bHzzIsDAQC}. As the matrix $\mathsf{M}_{\alpha\beta}$ is invertible, Rouch\'e-Frobenius theorem ensures that the solution is unique. It is noteworthy to notice that some of the times $t_\alpha$ might be negative. This means that those analog evolutions should be done with inverted coupling signs. However, there is actually a simple method to address this problem consisting in evolving with times $\tilde{t}_\alpha=t_\alpha+|t_{\mathrm{min}}|$ (see Appendix \ref{AppSec:Negative_times} for further details).

To sum up, in this section, we have designed an optimal DAQC protocol to construct an arbitrary inhomogeneous ATA two-body Ising Hamiltonian using as a resource an homogeneous ATA two-body Ising Hamiltonian and $x$-rotations. This protocol, which is quadratic in the total number of qubits, is optimal for a generic Hamiltonian, since it makes use of the same number of resources as degrees of freedom the Hamiltonian has. For the case of NN Hamiltonian, the protocol is even simpler, and requires only $N-1$ SQRs, i.e., one per site. These two protocols show, as a by-product, that the Ising model is universal, since it can be used to construct a ZZ gate between two arbitrary qubits.

\subsection{XZ model}\label{subsec:XZ_model}
As the Ising model is a universal Hamiltonian within the DAQC paradigm, we can simulate any other Hamiltonian evolution. In order to do it in a systematic manner, we will make use of a Trotter decomposition on top of the previous algorithm \cite{Lloyd_1996}. We illustrate this idea by explicitly constructing an inhomogeneous general two-body XZ Hamiltonian.

The unitary that we want to simulate is $U_{XZ}=e^{i t_F H_{XZ}}$, with $H_{XZ}= \sum_{j<k}^N \sum_{\mu,\nu=\{x,z\}}g_{jk}^{\mu\nu}\sigma_\mu^j \sigma_\nu^k$. We firstly perform a Trotter decomposition $U_{XZ}\approx (e^{i \frac{t_F}{n_T} H_{XZ}})^{n_T}$, with $n_T$ the number of Trotter steps,
\begin{eqnarray}
H_{XZ}&=&  \sum_{j<k}^N \sum_{\mu,\nu}g_{jk}^{\mu\nu}\sigma_\mu^j \sigma_\nu^k= \sum_{j<k}^N \sum_{\mu,\nu,s}g_{jk}^{(s)}\alpha_j^{(\mu,s)}\alpha_k^{(\nu,s)}\sigma_\mu^j \sigma_\nu^k\label{eq:H_I_XZ}\nonumber\\
&=&\sum_{j<k}^N \sum_{s=1}^4 g_{jk}^{(s)}\left(\cos(\theta_j^{(s)})\sigma_z^j+\sin(\theta_j^{(s)})\sigma_x^j\right)\times\nonumber\\
&&\left(\cos(\theta_k^{(s)})\sigma_z^k+\sin(\theta_k^{(s)})\sigma_x^k\right),
\end{eqnarray}
with $s$ running from $\{1,\dots, 4\}$, which are the number of combinations of the two types of couplings (for instance, for the XYZ Hamiltonian, it would run from $1$ to $9$). The implicit definition of parameters is $\alpha_j^{(x,s)}=\sin(\theta_j^{(s)})$ and $\alpha_j^{(z,s)}=\cos(\theta_j^{(s)})$. We decompose the pairs of operators with their coefficients in homogeneous $\sigma_z^j\sigma_z^k$ operators with local rotations $R_{\theta_j^{(s)}}=\left(\cos(\theta_j^{(s)}/2)\sigma_z^j+\sin(\theta_j^{(s)}/2)\sigma_x^j\right)$,
\begin{equation}
R_{\theta_j^{(s)}} \sigma_z^j R_{\theta_j^{(s)}}=\left(\cos(\theta_j^{(s)})\sigma_z^j+\sin(\theta_j^{(s)})\sigma_x^j\right),\nonumber
\end{equation}
for all pairs of qubits. This rotation is produced by the Hamiltonian $H_{\theta_j^{(s)}}=(\pi/2)\left(-\mathbbm{1}+\cos(\theta_j^{(s)})\sigma_z^j+\sin(\theta_j^{(s)})\sigma_x^j\right)$.  We have to perform the SQRs in all qubits $R_{\theta^{(s)}}=\otimes_w^N R_{\theta_w^{(s)}}$. The total Hamiltonian is reconstructed as 
\begin{equation}
H_{XZ}=\sum_{s=1}^4 R_{\theta^{(s)}}\left(g_{jk}^{(s)}\sigma_z^j \sigma_z^k\right) R_{\theta^{(s)}}^\dagger.
\end{equation}
This equality is only valid in the Hamiltonian and not in the total unitary evolution. In each Trotter time step $t_T=\left(t_F/n_T\right)$ we must evolve according to $H_{ZZ}^{(s)}=g_{jk}^{(s)}\sigma_z^j \sigma_z^k$ between a pair of sets of SQRs $R_{\theta^{(s)}}$
\begin{eqnarray}
U_{XZ}\approx\left(\prod_{s=1}^4 R_{\theta^{(s)}}e^{i\frac{t_T}{n_T}(H_{ZZ})^{(s)}} R_{\theta^{(s)}}\right)^{n_T}.
\end{eqnarray}
The problem is reduced to find a set of phases $\theta_w^{(s)}$ such that the system of equations $g_{jk}^{\mu\nu}=g_{jk}^{(s)}\alpha_j^{(\mu,s)}\alpha_k^{(\nu,s)}$ has independent solutions. A specific set of phases which adequately plays this role is $\theta_w^{(s)}=\frac{s\pi w}{2(w+1)}$, with distance between two nearest-neighbour qubit phases scaling polynomially $d(s,w)=\frac{s\pi}{2(w^2+3w+2)}$. See Appendix \ref{Appsec:XZmodel_couplings} for further details. 

To sum up, in this section, we have proposed a protocol to construct an arbitrary two-body Hamiltonian by exploiting the freedom in the phases of the SQRs and the Trotterization technique. This protocol requires $4N(N-1)$ ($4(N-1)$) analog blocks per Trotter step for the ATA (NN) XZ model in total, which is optimal for a generic two-body Hamiltonian in terms of the free coefficients to fix. Similarly, it can be proven that a trivial extension to the general two-body $XYZ$ model requires more angles per rotation summing up to $9N(N-1)$ ($9(N-1)$) blocks per Trotter step for the ATA (NN) model. 
\begin{figure}[h]
	\includegraphics[width=1\linewidth]{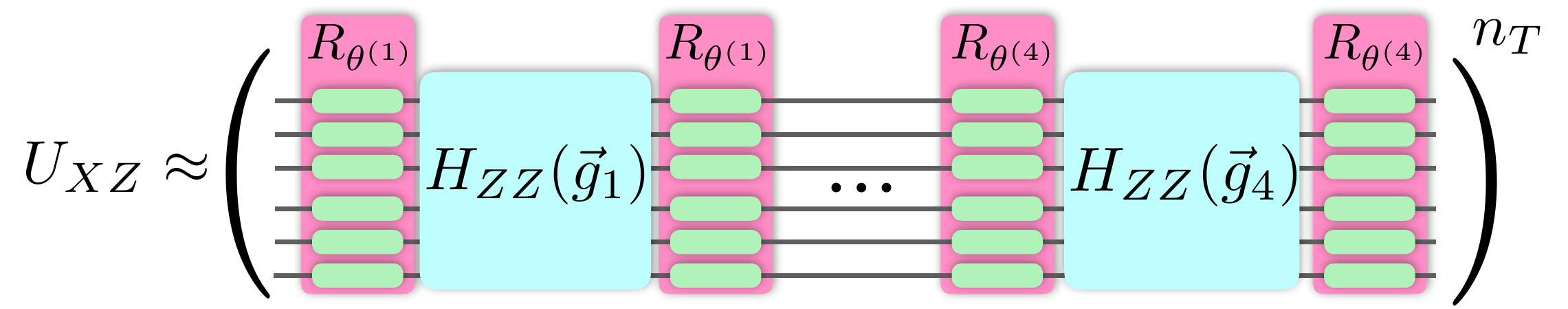}
	\caption{\label{fig:HzxI_HzzI_sDAQC} \textbf{Algorithm to simulate the inhomogeneous XZ model from a Ising model}. The combination of four different general Ising evolutions and rotations handles the minimum degrees of freedom for each term $g_{jk}^{\mu\nu}$. $R_{\theta^{(s)}}$ rotations can be combined with inner $\sigma_x^n \sigma_x^m$ rotations  required to implement the inhomogeneous ZZ Ising model into unique blocks of SQRs.}
\end{figure}

\subsection{$\mathbf{M}$-body Hamiltonian}\label{Subsec:M_body}

With similar techniques, we can systematically construct the evolution of a completely general Hamiltonian with up to $M$-body interactions. For the sake of clarity, we sketch here the sequence of steps to simulate an arbitrary $4$-body NN Hamiltonian, explaining how to generalize it to $M$-body Hamiltonians in the end of the work. We have chosen $M=4$ to illustrate the protocol, since it is a case of especial interest in quantum chemistry and nuclear physics,
\begin{eqnarray}
	H_{4b}&=&\sum_{j \chi\eta} g_{(2,j)}^{\chi\eta}\sigma_{\chi}^{j}\sigma_{\eta}^{j+1}+\sum_{j\chi\eta\gamma} g_{(3,j)}^{\chi\eta\gamma}\sigma_{\chi}^{j}\sigma_{\eta}^{j+1}\sigma_{\gamma}^{j+2}\nonumber\\
	&&+\sum_{j \chi\eta\gamma\rho}g_{(4,j)}^{\chi\eta\gamma\rho}\sigma_{\chi}^{j}\sigma_{\eta}^{j+1}\sigma_{\gamma}^{j+2}\sigma_{\rho}^{j+3},
\end{eqnarray}
where the indices run over $\{\chi,\eta,\gamma,\rho\}\in\{x,y,z\}$ and $j \in\{1,...,N\}$. In the first stage, an inhomogeneous ZZ-Ising Hamiltonian is sandwitched between rotated XX-Ising evolutions and their conjugate transposes, 
\begin{eqnarray}
	H_1^{(k)}=e^{-iO_{XX}^{1(k)}}H_{ZZ}^{1(k)}e^{iO_{XX}^{1(k)}},\label{eq:Mbody_H1k}\\
	H_2^{(k)}=e^{-iO_{XX}^{2(k)}}H_{ZZ}^{2(k)}e^{iO_{XX}^{2(k)}},\label{eq:Mbody_H2k}
\end{eqnarray}
with $O_{XX}^{1(k)}=\Phi^{(k)}_1\sigma_x^{1}\sigma_x^{2}+\Phi^{(k)}_3\sigma_x^{3}\sigma_x^{4}+\Phi^{(k)}_5\sigma_x^{5}\sigma_x^{6}+...$ and its translationally shifted $O_{XX}^{2(k)}=\Phi^{(k)}_2\sigma_x^{2}\sigma_x^{3}+\Phi^{(k)}_4\sigma_x^{4}\sigma_x^{5}+\Phi^{(k)}_6\sigma_x^{6}\sigma_x^{7}+...$ are built from evolutions of ZZ-Ising models rotated with single qubit gates in all qubits $R=\otimes_j\left(\cos(\pi/4)\sigma_z^j+\sin(\pi/4)\sigma_x^j\right)$. We would like to remark that operators $O_{XX}^{1(k)}$ and $O_{XX}^{2(k)}$ contain interactions separated by the interaction length $L=M/2=2$, e.g.,  $O_{XX}^{1(k)}$ has a term $\sigma_x^{1}\sigma_x^{2}$, but not $\sigma_x^{2}\sigma_x^{3}$. In order to simulate, for example, a three-body (five-body) Hamiltonian, we would need a different decomposition with 3 (5) translationally invariant sets of blocks (see further details in Appendix \ref{Appsec:Mbody}). 

It is straightforward to see that Hamiltonian $H_1^{(k)}$ contains all two-body and three-body terms with different supports but not all four-body terms, i.e.,  $H_1^{(k)}=h_{12}+h_{23}+...+{h_{N-1,N}}+h_{123}+h_{234}+...+{h_{N-2,N-1,N}}+h_{1234}+h_{3456}+h_{5678}+...$, where $h_{ij...}$ is an operator acting non-trivially on qubits $\{i,j,...\}$. On the other hand, $H_2^{(k)}$ contains (again) all two-body and three-body terms, as well as the complementary four-body terms $H_2^{(k)}=...+h_{2345}+h_{4567}+h_{6789}+...$.

The coefficients of operators $h_{ij...}$ are coupled together. For the simulation of an arbitrary four-body NN Hamiltonian with the requested support, it is sufficient to sum four of each of the blocks $H_0=\sum_{k=1}^4 H_1^{(k)}+H_2^{(k)}$ to disentangle the parameters, generating at least one term operating in each support. Finally, to create all XYZ operators, we need to concatenate $3^{M}=81$ $H_0$-blocks interleaved by single qubit rotations with the estructure $R^{(l)}=\otimes_j^N r_j^{(l)}\sigma_x^{j}+s_j^{(l)}\sigma_y^{j}+t_j^{(l)}\sigma_z^{j}$, i.e., 
\begin{equation}
	H_{4{b}}=\sum_l R^{(l)} H_0^{(l)} R^{(l)},
\end{equation}
with $(r_j^{(l)},s_j^{(l)},t_j^{(l)})$ fulfilling the constraint $|r_j^{(l)}|^2+|s_j^{(l)}|^2+|t_j^{(l)}|^2=1$. The aforementioned construction works for one Trotter step, so it must be repeated $n_T$ times to approximate the evolution as $U_{4{b}}=e^{-iH_{4{b}}t}\approx (e^{-iH_{4{b}}t/n_T})^{n_T}$. The total number of analog blocks, engineered time slices, for the most general simulation with up to $M$-body interactions and $N$ number of qubits is $a(M)N+b(M)$ with, 
\begin{eqnarray}
	a(M)&=\frac{9}{4}\left(3^{M-1}-3\right),\\
	b(M)&=\frac{3^{M-1}}{2}\left(\frac{3}{2}-M\right).
\end{eqnarray}
In other words, it grows linearly with the number of qubits and exponentially with the number of body interactions. For the four-body system here described, the total number is $117N-306$. A more detailed explanation of the algorithm for simulating NN Hamiltonians with general $M$-body interactions can be found in Appendix \ref{Appsec:Mbody}.
\begin{figure}[h]
	\includegraphics[width=1\linewidth]{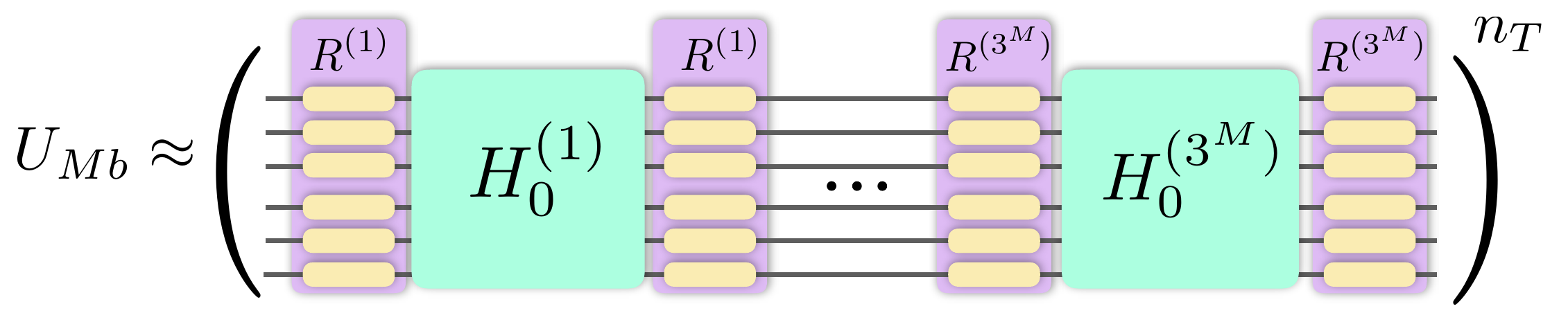}
	\caption{\label{fig:U_4b} \textbf{Algorithm to implement the $M$-body evolution.} Generalized rotations $R^{(l)}=\otimes_j^N r_j^{(l)}\sigma_x^{j}+s_j^{(l)}\sigma_y^{j}+t_j^{(l)}\sigma_z^{j}$ sandwich evolutions of $H_0^{(l)}=\sum_{k=1}^4 (H_1^{(k)})^{(l)}+(H_2^{(k)})^{(l)}$, see Eqs. (\ref{eq:Mbody_H1k}) and (\ref{eq:Mbody_H2k}) and Appendix \ref{Appsec:Mbody} for an example with $M=4$.}
\end{figure}

\section{Stepwise and Banged DAQC}
An important source of errors in realistic quantum algorithms comes from turning on and off multi-qubit quantum gates. In the interest of reducing its effect, we introduce the concept of banged digital-analog quantum computing (bDAQC) as a different way to perform quantum algorithms, in opposition to the stepwise digital-analog quantum computing (sDAQC) previously introduced. 

The term {\it bang-bang} has been routinely used in classical control theory \cite{ControlTheory_Bangbang,ControlTheory_Bangbang2} and was first introduced in the field of quantum physics as a tool for dynamically decoupling \cite{Viola_1998_DynamicalDec,Morton_2006_BangBang} a controlled quantum system from its environment. Here, however, we use the term {\it banged} to express the fact that the analog evolution is not switched off while the fast SQRs pulses are being turned on. Naturally, the bDAQC introduces an additional digital error in contrast with the sDAQC that can be, in the best case, of third order in time. We argue that this error competes with the experimental error produced by the switching of multi-qubit gates and depending on the cases might prove less harmful to the algorithm.

\begin{figure}[h]
	\includegraphics[width=1\linewidth]{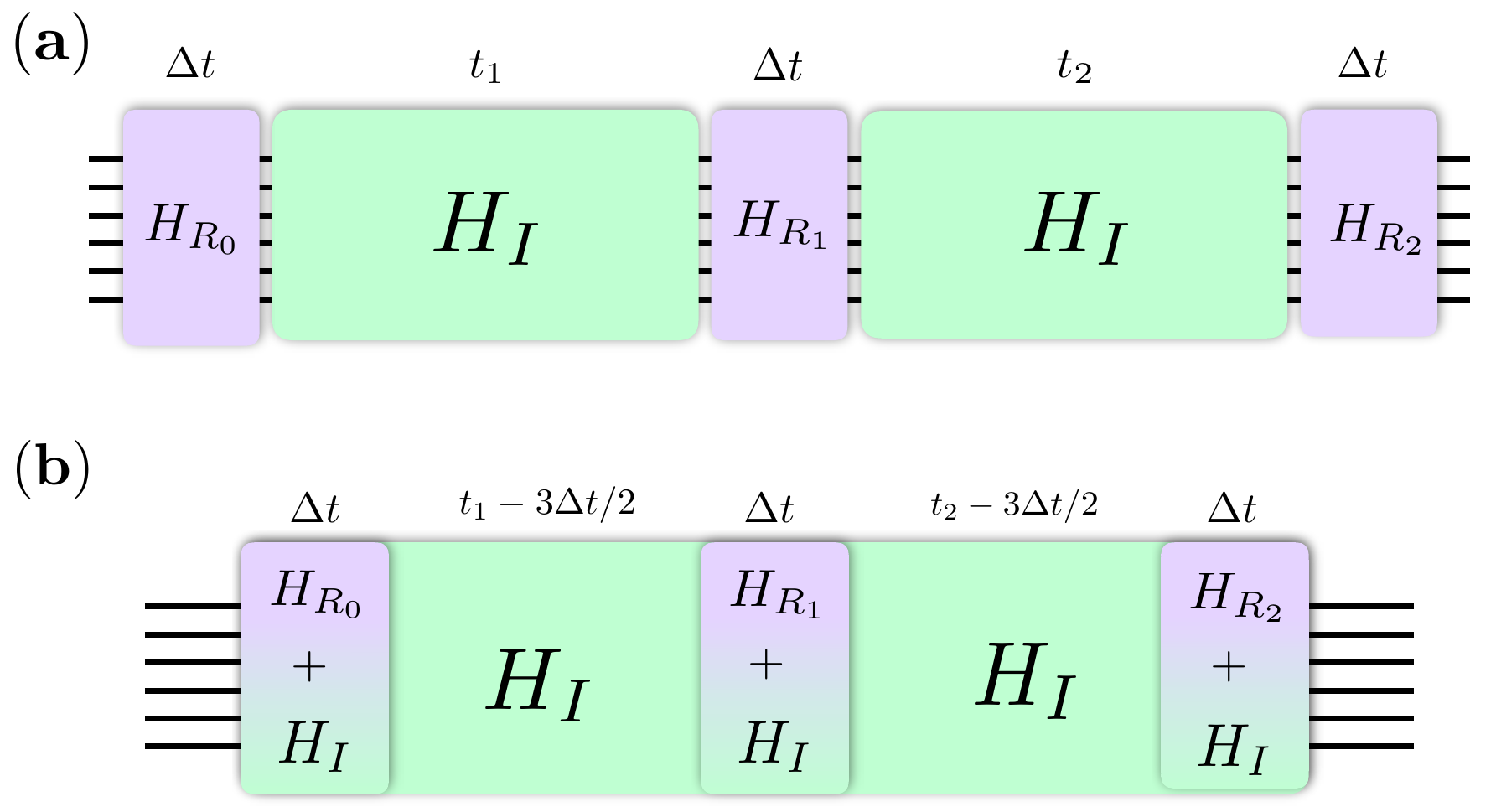}
	\caption{\label{fig:sDAQC_vs_bDAQC} \textbf{Comparison between a sDAQC and bDAQC protocol.} (a) sDAQC: digital evolutions with Hamiltonian $H_{R_k}$, and analog blocks ones evolving with Hamiltonians $H_I$ are well separated in time under the sudden approximation. (b) bDAQC: the adiabatic evolution $H_I$ is on during the whole time and fast rotations are added ($H_{R_k}+H_I$), also under the sudden approximation.}
\end{figure}

\subsection{Stepwise DAQC}
In a typical sDAQC scenario, a total unitary evolution is built interleaving an evolution of a fixed entangling Hamiltonian $H_I$ with sets of SQRs as
\begin{equation}
U_T = ...U_{R_2} e^{-i H_I t_2} U_{R_1}e^{-i H_I t_1}U_{R_0},\label{eq:sDAQC_UT}
\end{equation}
with $U_{R_n}$ a general digital rotation operator acting on any subset of the Hilbert space (e.g. in two qubits), see Fig. \ref{fig:sDAQC_vs_bDAQC}(a). The sudden approximation states that we can implement such evolution $U(t)\approx e^{-it_FH(t)}$ with a time dependent Hamiltonian 
\begin{eqnarray}
H(t)=& \sum_{n=0}H_{R_n} \Pi_{\Delta t}(t-[T_n+n\Delta t]) \nonumber\\
&+ H_I \Pi_{T_{n+1}}(t-[T_{n}+(n+1)\Delta t]),\label{eq:H_time_PDAQC}
\end{eqnarray}
where $H_{R_n}$ gives rise to the digital unitary evolution $U_{R_n}=e^{-iH_{R_n}\Delta t}$. We have defined $T_n=\sum_{r=0}^{n}t_{r}$, assuming $t_0=0$, and the rectangular window function 
\begin{equation}
\Pi_{T}(t-t_s)= \theta(t-t_s)-\theta(t-(t_s+T)),
\end{equation}
with $\theta(t)=1$ for $t\ge0$ and 0 otherwise. 

\subsection{Banged DAQC}
The aforementioned stepwise protocol assumes that we can turn on and off the Hamiltonians $H_I$ and $H_{R_n}$ infinitely fast; something obviously unphysical. The bDAQC protocol consists on implementing the same evolution without turning off the background Hamiltonian $H_I$ and performing short (and intense) pulses to implement the single qubit rotations. 

Exploiting the symmetrized exponential decomposition for all the SQRs blocks, the ideal evolution of the system in Eq. (\ref{eq:sDAQC_UT}) can be performed without turning off the entangling Hamiltonian
\begin{eqnarray}
H(t)&= H_I+\sum_n^{N-1} H_{R_n}\Pi_{\Delta t}(t-[T_n-\Delta t/2])\nonumber\\
&+ H_{R_N}\Pi_{\Delta t}(t-[T_N-\Delta t]),\label{eq:H_time_bDAQC}
\end{eqnarray}
where again $T_n=\sum_{r=0}^{n}t_{r}$, see Fig. \ref{fig:sDAQC_vs_bDAQC}(b).

\subsection{Error estimation}
The additional error per step that we are introducing can be estimated with the aid of a Schatten norm as the difference between evolving the system with the sDAQC protocol with respect to the bDAQC
\begin{eqnarray}
e_n &=& ||1-e^{-i H_{I} \Delta t/2}e^{-i H_{R_n} \Delta t}e^{-i H_{I} \Delta t/2}e^{i (H_I+H_{R_n}) \Delta t} ||\nonumber\\
&=&\frac{(\Delta t)^3}{4}||[[H_I,H_{R_n}],H_I+2H_{R_n}]||+O((\Delta t)^4),
\end{eqnarray}
where we have made use of Zassenhaus formula. The first and last blocks at the boundaries introduce second order errors $e_{0,N}=O((\Delta t)^2)$, as the evolutions cannot be symmetrized. The total digital error of the banged protocol is the sum $E=\sum_n e_n = A e_n$, with $A\propto O(N)$ or $O(N^2)$ given that the total number of analog blocks increases polinomially (linearly or quadratically) with the total number of qubits $N$ for NN or ATA coupling configurations respectively. 

\section{Example: XZ model}
Let us exemplify this new paradigm of protocols with a numerical simulation of the non-commuting ATA XZ model previously described in Subsec. \ref{subsec:XZ_model}. We first compare the exact performance of a purely digital protocol (DQC) with both the sDAQC and bDAQC.  We decompose at each Trotter step, i.e.,  $U(t_F)\approx U(t_T)^{n_T}$, $t_T=t_F/n_T$, the evolution of the terms in the ATA DQC protocol
\begin{equation}
	e^{i \varphi_{jk}^{\mu\nu}\sigma_{\mu}^j\sigma_{\nu}^k}=e^{i \frac{\pi}{4}\sigma_{y}^j}e^{i \frac{\pi}{4}\sigma_{\mu}^j\sigma_{\nu}^k}e^{i\varphi_{jk}^{\mu\nu}\sigma_{y}^j}e^{i \frac{\pi}{4}\sigma_{\mu}^j\sigma_{\nu}^k}e^{-i \frac{\pi}{4}\sigma_{y}^j}
\end{equation}
where $\varphi_{jk}^{\mu\nu}=t_T g_{jk}^{\mu\nu}$ into fixed $\pi/4$ two-qubit gates. Such evolution might be implemented directly for qubits with strong coupling, typically neighbours. However, such evolution could be decomposed into NN protocols, e.g.,  with swap gates; protocols that typically improve the performance. Another option for complex systems is the use genetic algorithms to optimize this decomposition \cite{Las_Heras_2016}.
\begin{figure}[h] 
	\includegraphics[width=.95\linewidth]{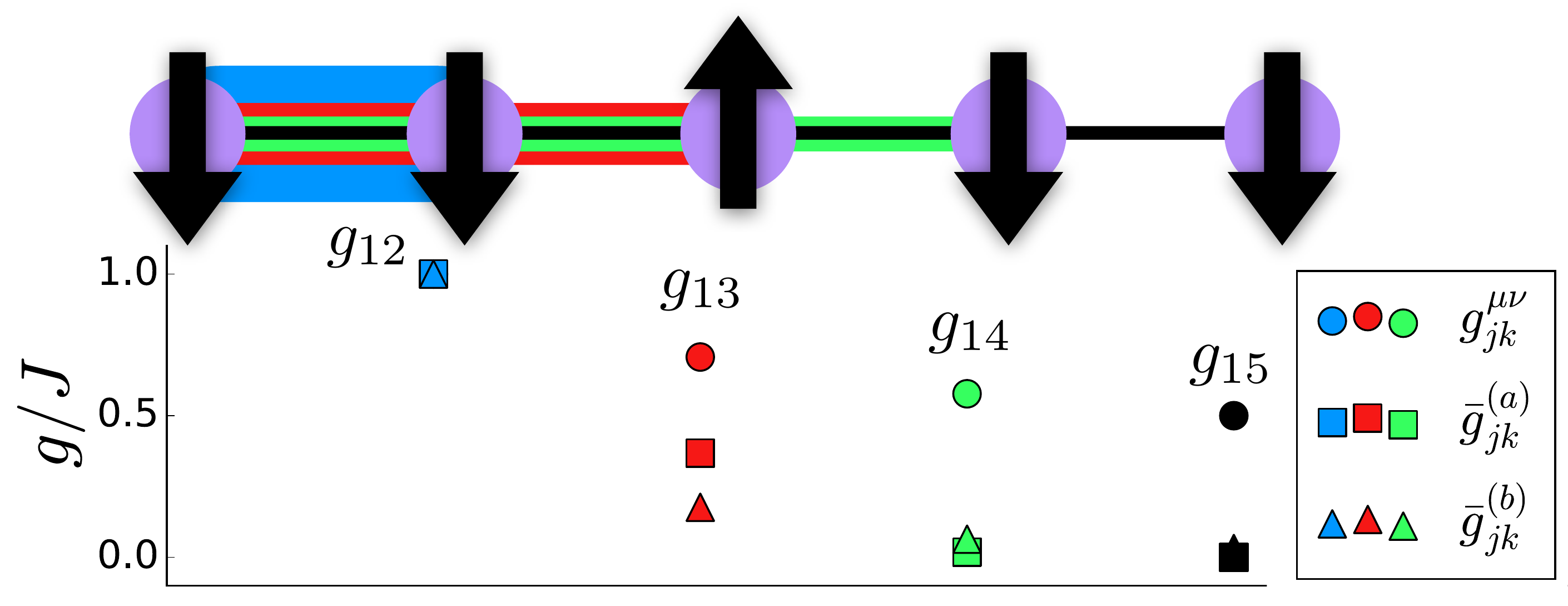}
	\caption{\label{fig:XZ_sim_coup} \textbf{5 qubit spin system.} We have simulated the XZ model with coupling parameters of the original system Hamiltonian with polynomial $\bar{g}_{jk}^{a}=J/|j-k|^{5/2}$ (squares), and exponential $\bar{g}_{jk}^{b}=J e^{-(|j-k|-1)^2}$ (triangles) decay, corresponding to the simulations in Fig. \ref{fig:XZ_sim_fidelity_DigError}. On the other hand, the couplings of the simulated one are $g_{jk}^{\mu\nu}=J/|j-k|^{1/2}$ (circles); $\forall\,j<k$ and $j,k\in\{1-5\}$ and $\forall\mu,\nu\in\{x,z\}$.}
\end{figure}

\begin{figure}[h] 
	\includegraphics[width=1\linewidth]{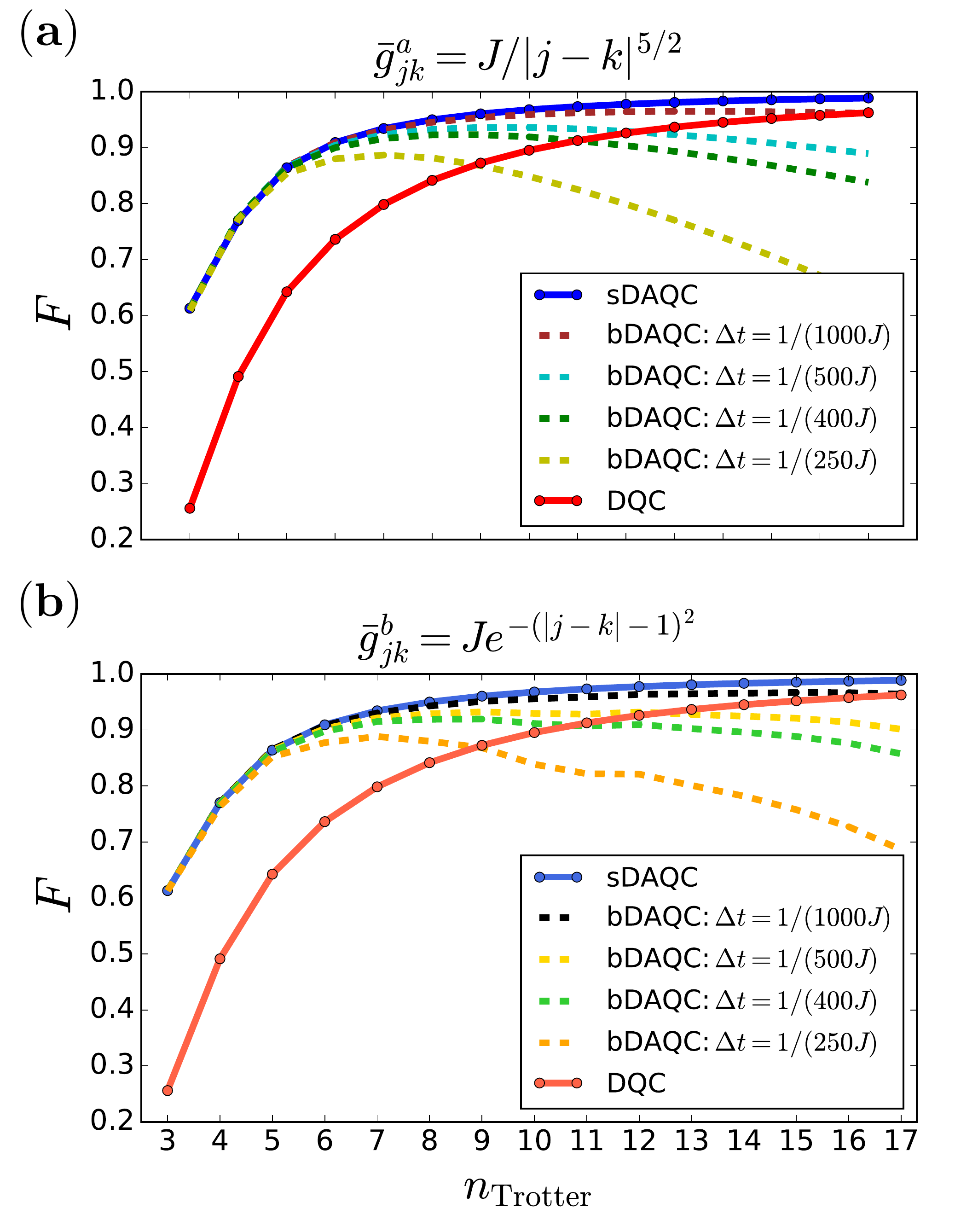}
	\caption{\label{fig:XZ_sim_fidelity_DigError} \textbf{Exact Trotter evolution: sDAQC vs bDAQC vs DQC.} Fidelity $F=|\langle \Psi_{e}|\Psi_{c}\rangle|^2$  between exact and ideal  DQC, sDAQC and bDAQC protocols for an evolution of a 5-qubit XZ model as a function of Trotter steps for a final time $t_F=1/J=2$. The couplings of the original system Hamiltonian are (a) polynomially and (b) exponentially decaying; see Fig. \ref{fig:XZ_sim_coup}. The initial state is $\ket{\Psi(0)}=\ket{\downarrow\downarrow\uparrow\downarrow\downarrow}$. The bDAQC protocol considers different finite times $\Delta t$ for performing the SQRs blocks where the entangling Hamiltonian $H_I=\bar{H}_{ZZ}$ is still on.}
\end{figure}

We have used two different fixed entangling ATA ZZ Hamiltonians $\bar{H}_{ZZ}=\sum_{j<k}\bar{g}_{jk}\sigma_z^j\sigma_z^k$ with  (a)  $\bar{g}_{jk}=J/|j-k|^{5/2}$ and (b) $\bar{g}_{jk}=Je^{-(|j-k|-1)^2}$ parameter couplings to simulate Hamiltonian (\ref{eq:H_I_XZ}) with $g_{jk}^{\mu\nu}=J/|j-k|^{1/2}$ couplings where $\forall\mu,\nu\in\{x,z\}$ and $\forall\,j<k$ and $j,k\in\{1-5\}$, see Fig. \ref{fig:XZ_sim_coup}. In Fig. \ref{fig:XZ_sim_fidelity_DigError}, it is plotted the fidelity as a function of Trotter steps between exact and computed states $F=|\langle \Psi_{e}|\Psi_{c}\rangle|^2$ of the exact evolution of a 5-qubit XZ model for a final time $t_F=1/J=2$; where the initial state has an excitation in the 3 qubit, $\ket{\Psi(0)}=\ket{\downarrow\downarrow\uparrow\downarrow\downarrow}$. It is clear that fidelities above $90\%$ can be achieved even for the bDAQC protocol taking into consideration different finite times $\Delta t$ for performing the SQRs blocks where the $H_I=\bar{H}_{ZZ}$ is still on, and using two different base entangling Hamiltonians.
\subsection{Computational times}
\label{subsec:computational_times}
The times $t_{\alpha}^{(s)}=t_T \mathsf{M}_{\alpha\beta}^{-1}(\bar{g}_{\beta\beta}^{(s)})^{-1}g_{\beta}$ of the analog blocks required to perform the four ZZ-DAQC protocols (both stepwise or banged) might be negative. Here, $\bar{g}_{\beta\beta}^{(s)}$ is defined as the diagonal matrix with elements $\bar{g}_{jk}^{(s)}$ in the same $\beta(j,k)$ ordering. In such case, there is a simple way around the problem of evolving with negative times (or inverted coupling signs) by realizing that the constant vector is an eigenvector of the $\mathsf{M}_{\alpha\beta}$ matrix, see demonstration in Appendix \ref{AppSec:Negative_times}. 
\begin{figure}[h]
	\includegraphics[width=1\linewidth]{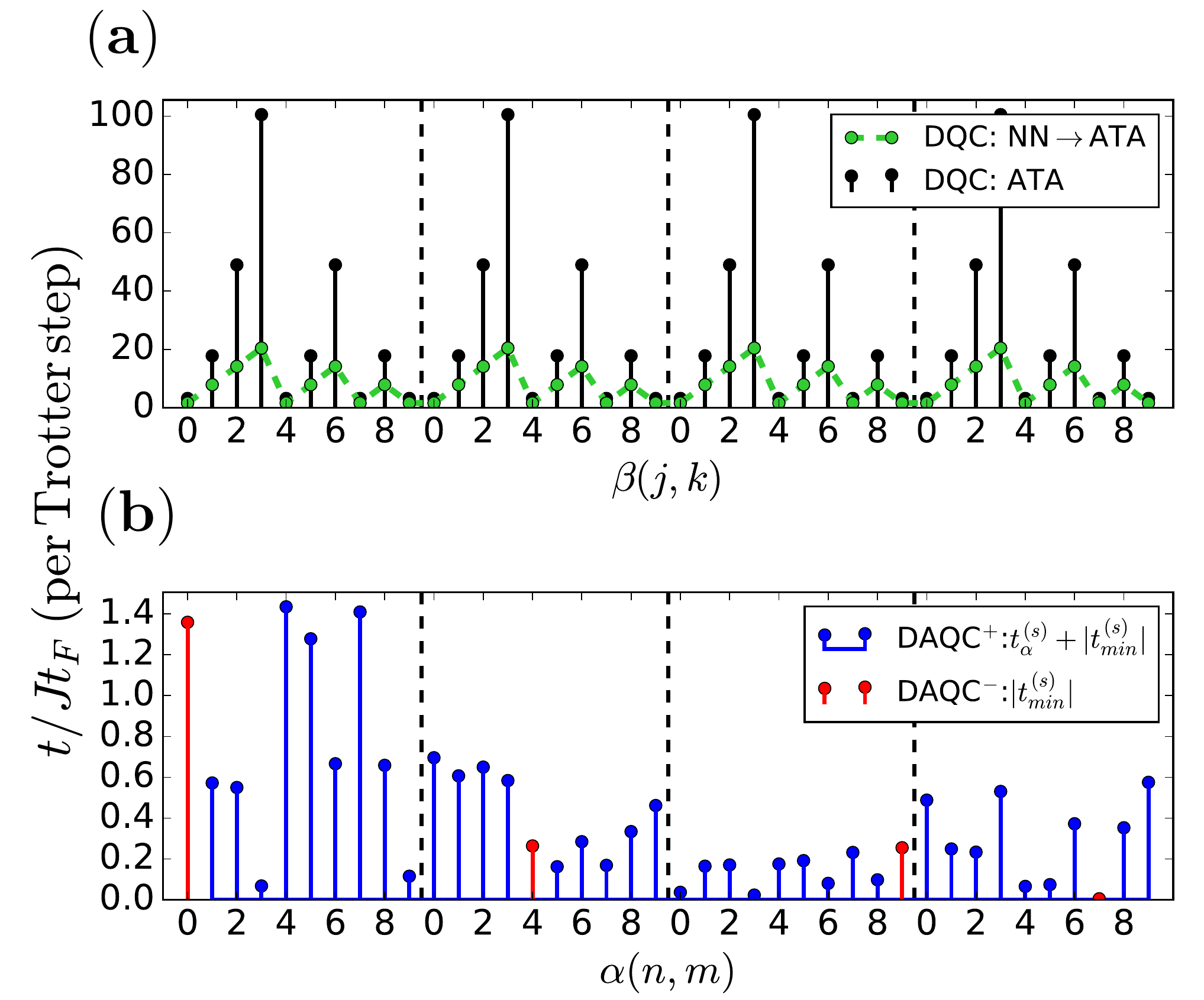}
	\caption{\label{fig:XZ_sim_times} \textbf{Times in a Trotter step to perform the DQC and DAQC protocols in terms of the simulated two-body interaction term index $\beta$ and the analog block index $\alpha$ respectively for the simulation in Fig. \ref{fig:XZ_sim_fidelity_DigError}(a).} (a) (Black) ATA DQC protocol improves considerably time resources when it is decomposed into NN SWAP gates (green). (b) However, both of them still perform worse than the DAQC protocol; implemented through 4 ZZ-blocks of times $t_\alpha^{(s)}$ (blue points), in which the most negative times $|t_{\mathrm{min}}^{(s)}|$ have been highlighted (red points).}
\end{figure}

\begin{figure}[h]
		\includegraphics[width=.95\linewidth]{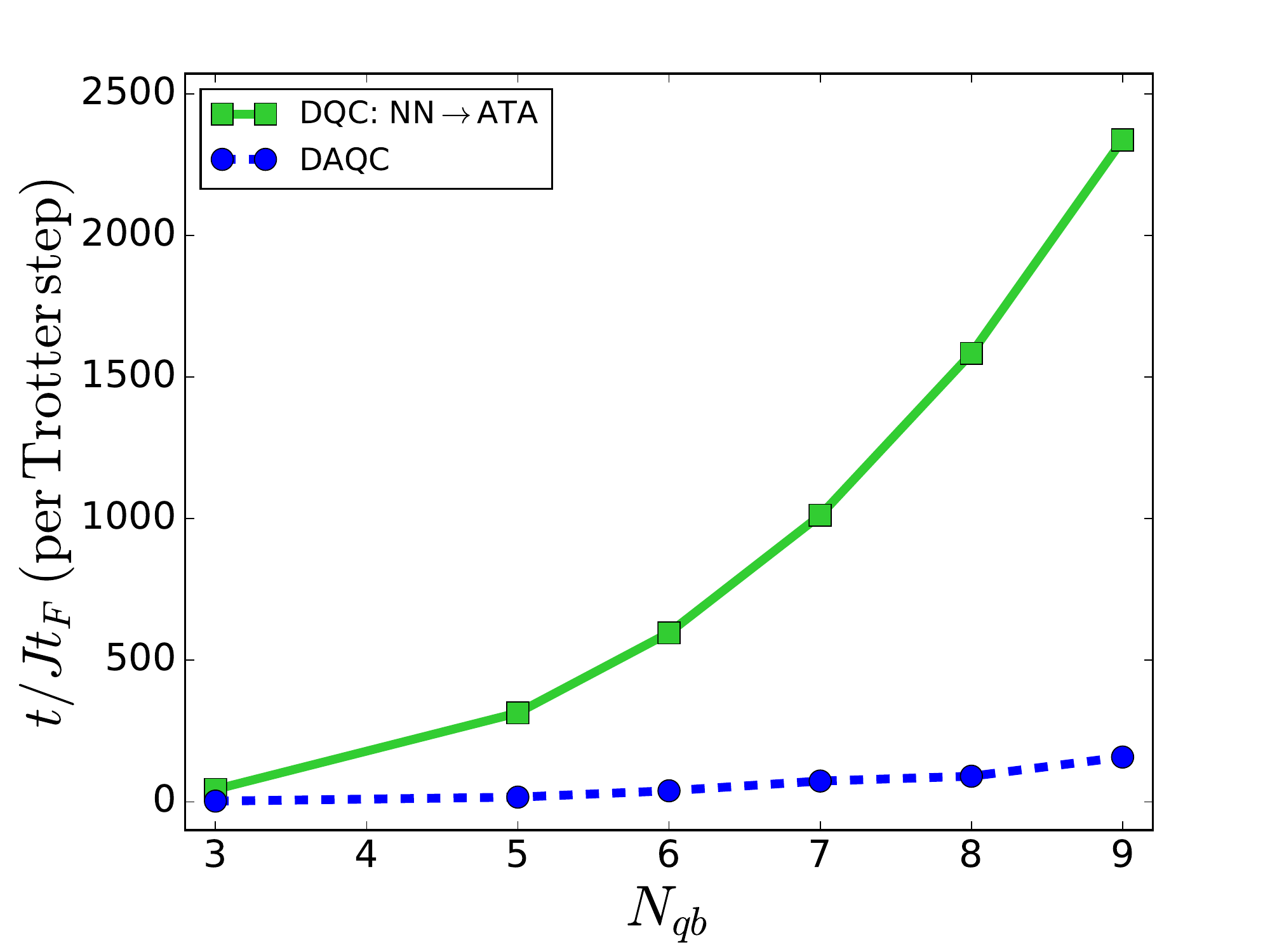}
		\caption{\label{fig:XZ_sim_total_times} \textbf{Total times (sums of points in Fig. \ref{fig:XZ_sim_times}) for each Trotter step as a function of number of qubits.} The (blue) DAQC protocol requires (polinomially with $N$) less time than the (green) DQC protocol as the number of qubits increases. This systematic protocol used is not valid for $N=4$ qubits.}
\end{figure}

In Fig. \ref{fig:XZ_sim_times}(a), we show the times of the two qubit gates in the DQC protocol in a Trotter step by directly using the ATA $\bar{H}_{ZZ}$ Hamiltonian or the most efficient protocol that decomposes each $\sigma_z^j\sigma_z^k$ interaction into NN fixed $\pi/4$ gates, i.e.,  the gate decomposition $U_{\mathrm{SWAP}}(j,j+1)=e^{i\pi/4\sigma_x^j\sigma_x^{j+1}}e^{i\pi/4\sigma_y^j\sigma_y^{j+1}}$. In virtue of simplicity, we have omitted the times required for performing SQRs to transform $\sigma_z^j\sigma_z^{j+1}$ gates into $\sigma_x^j\sigma_x^{j+1}$ or $\sigma_y^j\sigma_y^{j+1}$. In opposition, Fig. \ref{fig:XZ_sim_times}(b) shows the times of the analog blocks in a DAQC protocol, clearly smaller. Both Figs. \ref{fig:XZ_sim_times}(a) and (b) use the same parameters of the simulation in Fig. \ref{fig:XZ_sim_fidelity_DigError}(a). The total sum of times in a Trotter step of the analog blocks (DAQC) or $\pi/4$ two-qubit gates (optimised DQC) for same simulation parameters varying the number of qubits is sketched in Fig. \ref{fig:XZ_sim_total_times}. It becomes clear that the the loss of coherence and population is going to affect faster the evolutions decomposed in purely digital gates.
\subsection{Experimental error simulation}
\begin{figure}[h]
	\includegraphics[width=.98\linewidth]{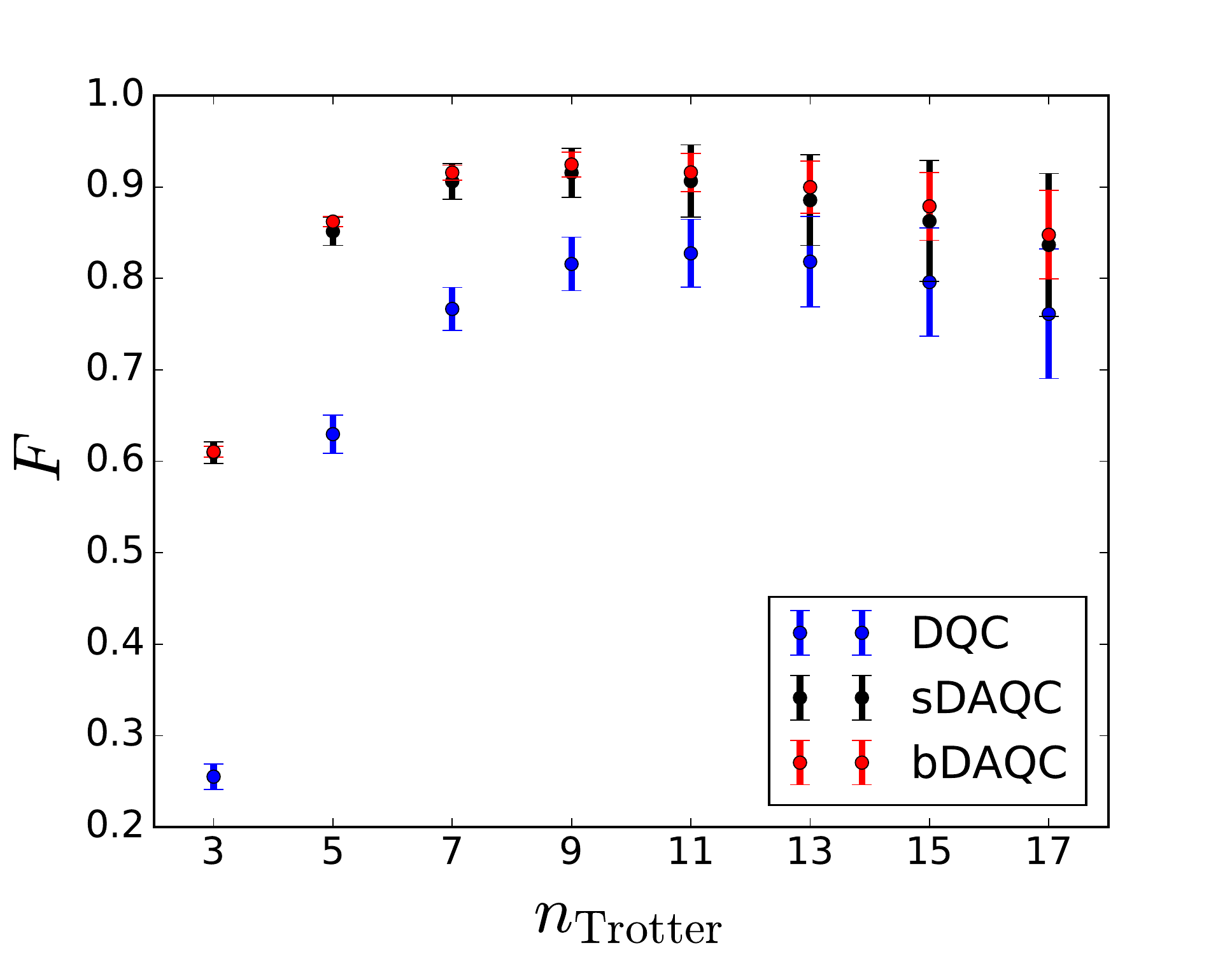}
	\caption{\label{fig:XZ_sim_fidelity_DigAndExpErrorS} \textbf{Trotter evolution with errors: sDAQC vs bDAQC vs DQC.} The DQC protocol (blue) is simulated with a Gaussian phase noise with deviation $\xi_{D}=\mathcal{N}(0,\sigma_{D})$ added to the $\pi/4$ phases, where $\sigma_D=0.009$. sDAQC (black) and bDAQC (red) have Gaussian time noise in the analog blocks $\delta_s=\mathcal{N}(0,r_s \Delta t)$ and $\delta_b=\mathcal{N}(0,r_b \Delta t)$ respectively, and $r_b=0.9=r_s/2$. All simulations include random axis magnetic field noise $\Delta B_\gamma=\mathcal{U}(-r_U \Delta t/2,r_U \Delta t/2)$, with $r_U=0.002$. Rest of parameters are equal to Fig. \ref{fig:XZ_sim_fidelity_DigError}(a) where $\Delta t=1/500J$ (orange line for bDAQC).}
\end{figure}

We have performed a stochastic simulation of presumably leading order dephasing errors to compare the DQC and DAQC protocols, using the same parameters of Fig. \ref{fig:XZ_sim_fidelity_DigError}(a). Results are depicted in Fig. \ref{fig:XZ_sim_fidelity_DigAndExpErrorS}. In the DQC protocol, we have included two kinds of errors in all two-qubit blocks, (i) a Gaussian phase noise with deviation $\xi_{D}=\mathcal{N}(0,\sigma_{D})$ added to the $\pi/4$ phases and (ii)  single qubit operators simulating a uniform magnetic field noise $\Delta B_\gamma=\mathcal{U}(-r_U \Delta t/2,r_U \Delta t/2)$. Here, $\mathcal{N}(x,y)$ refers to a Gaussian noise distribution with mean $x$ and deviation $y$ and $\mathcal{U}(a,b)$ refers to a uniform noise distribution with range boundaries $(a,b)$.  The deviation ratio $\sigma_D=0.009$, compatible with a two qubit gate fidelity greater than $99\%$. On the other hand, $r_U=0.002$, a much slower random axis phase noise. In Fig. \ref{fig:XZ_sim_fidelity_DigAndExpErrorS} we have chosen a $\Delta t=1/500J$, corresponding with orange line of Fig. \ref{fig:XZ_sim_fidelity_DigError} of the bDAQC. For example, an ideal gate transforms into
\begin{equation}
e^{i\frac{\pi}{4}\sigma_\alpha^j\sigma_\beta^{k}}\rightarrow e^{i\frac{\pi}{4}(1+\xi_{D})\sigma_\alpha^j\sigma_\beta^{k}+\sum_{j,\gamma=x,y,z}\Delta B_\gamma^j \sigma_\gamma^j}.
\end{equation}

The bDAQC errors include: (i) Gaussian coherent noise in analog block times $t_\alpha^{(s)}\rightarrow t_\alpha^{(s)}+\delta_b$, where $\delta_b=\mathcal{N}(0,r_b \Delta t)$ plus single qubit gate errors modeled as those in the DQC. We have chosen a deviation ratio $r_b=0.9$ of the $\Delta t$ time required to perform a SQR. The total evolution of an analog block in the bDAQC becomes 
\begin{equation}
e^{i\bar{g}_{jk}t_\alpha\sigma_z^j\sigma_z^{k}}\rightarrow e^{i\bar{g}_{jk}(t_\alpha^{(s)}+\delta_b)\sigma_z^j\sigma_z^{k}+\sum_{j,\gamma=x,y,z}\Delta B_\gamma^j \sigma_\gamma^j}.
\end{equation}
Finally, the sDAQC analog blocks are transformed analogously to those in the bDAQC but the two qubit Gaussian and coherent phase noise has a stronger value capturing the switching on-off errors expected in an experiment, i.e.,  $t_\alpha^{(s)}\rightarrow t_\alpha^{(s)}+\delta_s$, with $\delta_s=\mathcal{N}(0,r_s \Delta t)$ and $r_s=2r_b$. Stochastic evolutions of the fidelity averaged over 1000 runs show a competition between sDAQC and bDAQC methods, both of them clearly above the DQC protocol. We emphasize here that the DQC simulation assumes that one can perform directly each of the ATA terms, i.e.,  a NN decomposition would perform worse in terms of dephasing than blue line in Fig. \ref{fig:XZ_sim_fidelity_DigAndExpErrorS} as it would require increasingly linear number of two-qubit gates.
\section{Quantum Architectures}
The DAQC paradigm proposed in this article is intended to be implemented on NISQ \cite{Preskill_2018} architectures where digital computation based on quantum error correction will still be beyond reach for several years or even decades. Different physical architectures are currently being investigated to perform quantum processing tasks, of which, those based on superconducting circuits and trapped ions are leading in performance and have potentially the brightest future. As stated above, the main requirements that the physical implementation must meet is the (simultaneous) single-qubit addressing to perform random-phase rotations, and a global entangling dynamics. 

{\it Superconducting Qubits.---} One of the most successful implementations of qubits is based on superconducting circuits with Josephson junctions; non-linear systems that play the role of artificial atoms. Different configurations of junctions, e.g., grounded (transmon qubits \cite{Koch_2007}) or in loops (flux qubits \cite{Chiorescu_2003}), can be coupled directly, or indirectly through transmission line resonators, and effectively modeled with Ising Hamiltonians with transverse fields \cite{Wendin_2017}
\begin{equation}
	H_{\mathrm{SQ}}/\hbar\approx\sum_i \frac{\omega_i}{2} \sigma_z^i+ \sum_{ij}g_{ij} \sigma_x^i \sigma_x^j.
\end{equation}
Here $\omega_i$ are the energies of each qubit system and $g_{ij}$ are the coupling parameters. Extra single-qubit  drivings can be added by either coupling the qubits (i) magnetically (in SQUID configurations) to external feed lines for $\sigma_z$ rotations, or (ii) electrically to the transmission line resonators as effective terms of capacitive couplings for $\sigma_x$ and $\sigma_y$ rotations. Going to an interaction picture, one can turn on and off the local terms to apply the single-qubit rotations when required \cite{Lamata_2018_DAQS}, to match the requirements of the above Sec. \ref{Sec:DAQC}. 

{\it Trapped Ions.---} In the Lamb-Dicke regime, a trapped-ion chain is also modeled with the Ising Hamiltonian \cite{Porras_2004}. Additional single-qubit addressing can be implemented by passing laser light through splitters and acousto-optic modulators to focus on single ions and engineer a.c.-Stark shifts \cite{Debnath_2016}
\begin{equation}
H_{\mathrm{Ions}}/\hbar\approx \sum_i  (B_i + W_i(t))\sigma_z^i+ \sum_{ij}J_{ij} \sigma_x^i \sigma_x^j.
\end{equation}
The coupling term can be engineered to have a polynomial decay $J_{ij}=J_{\mathrm{max}}/|i-j|^\alpha$ with $0<\alpha<3$. In practice, however, the parameter is typically set between $0.5<\alpha<1.8$, for experimental issues regarding heating of motional harmonic modes and an increase of decoherence \cite{Zhang_2017,Maier_2019}. For example, in \cite{Maier_2019}, local static $B_i $ and dynamic $W_i(t)$ transverse terms were engineered. Other experimental schemes achieve simultaneous single-qubit addressing by putting the ions in magnetic field gradients and using global microwave radiation \cite{Johanning_2009,Wanga_2009,Weidt_2016}.

\section{Conclusions}
We have shown how to construct digital-analog quantum algorithms based on a combination of single qubit rotations and multi-qubit entangling dynamics. With this DAQC approach, we have explicitly provided protocols to construct generic Ising, two-body, and $M$-body Hamiltonians. Furthermore, we have analysed a banged approach, bDAQC, as an alternative possibility to the stepwise protocol, sDAQC. In this way, to improve the fidelity of the global quantum dynamics, one can reduce the effect of switching on/off the entangling evolution. Numerical simulations performed in a XZ model with 5~qubits suggest the advantage of DAQC over DQC protocols in terms of fidelity and computational time, both in ideal and realistic scenarios. This alternative paradigm paves the way for the implementation of useful quantum algorithms beyond the current state-of-the-art in the NISQ era.
\\
\section*{Acknowledgements}
We thank I. Arrazola and J. Casanova for very fruitful, and encouraging discussions. We thank Giancarlo Gatti,  Ana Martin,  and Takahiro Tsunoda for critical reading of the manuscript. The authors acknowledge support from Basque Government IT986-16, and Ph.D. Grant No. PRE-2016-1-0284, Spanish MINECO/FEDER FIS2015-69983-P, Ram\'on y Cajal Grant RYC-2012-11391, the projects QMiCS (820505) and OpenSuperQ (820363) of the EU Flagship on Quantum Technologies, and EU FET Open Grant Quromorphic. This material is also based upon work supported by the U.S. Department of Energy, Office of Science, Office of Advance Scientific Computing Research (ASCR), Quantum Algorithms Teams project under field work proposal ERKJ335.

\appendix
\section{\uppercase{Negative times in the Ising model}}
\label{AppSec:Negative_times}

Some of the times $t_\alpha= \mathsf{M}_{\alpha\beta}^{-1}g_\beta(t_F/g)$ of the analog blocks to implement the Ising model can be negative. In a sDAQC protocol, one solution would be to do such analog evolutions with inverted coupling signs. However, in the bDAQC paradigm, we must keep untouched the base entangling Hamiltonian $\bar{H}_{ZZ}$. There is a simple way to mimic such behaviour by using a tantamount set of times that produces the same evolution. 

Given that the vector of times $t_{\alpha}=(t_1,t_2,...,t_{\mathrm{min}},...)$ with $t_{\mathrm{min}}<0$ and $t_{\mathrm{min}}< t_\alpha\,\forall \alpha$, there is an equivalent evolution with $N(N-1)/2-1$ blocks of times $\tilde{t}_{\alpha}=t_{\alpha}+|t_{\mathrm{min}}|e_1$ and an extra analog block with time $|t_{\mathrm{min}}|\lambda_1$, where $\lambda_1$ is the eigenvalue of a matrix $\mathsf{M}_{\alpha\beta}$ with corresponding constant eigenvector $e_1=\gamma(1,1,1...)^T$.

Let us first focus on the properties of the matrix $\mathsf{M}_{\alpha\beta}$ defined in Eq. (\ref{eq:Ising_Malphabeta}), created from the signs of applying sequentially gates $\sigma_x^n\sigma_x^m$ before and after the analog block evolutions. This matrix has an eigenvector $e_1=\gamma(1,1,1...)^T$ which corresponds to the eigenvalue $\lambda_1=N(N-9)/2+8$, that is negative for $N\in\{3,5,6\}$, and positive thereafter $N\in \mathbb{Z}\ge7$. Notice that we omit the case for $N=4$ as the matrix is non-invertible and different set of gates must be performed.

For $N<7$ we observe the following identity
\begin{eqnarray}
	&\mathsf{M}_{\alpha\beta}\left(t_{\alpha}+|t_{\mathrm{min}}|e_1-|t_{\mathrm{min}}|e_1\right)=\nonumber\\
	&=\mathsf{M}_{\alpha\beta}\left(t_{\alpha}+|t_{\mathrm{min}}| e_1\right)-\lambda_1|t_{\mathrm{min}}|e_1\nonumber\\
	&=\mathsf{M}_{\alpha\beta}\tilde{t}_\alpha+|\lambda_1t_{\mathrm{min}}|e_1
\end{eqnarray}
which corresponds to an evolution with $N(N-1)/2-1$ blocks of times $\tilde{t}_\alpha$.  One block has zero time ($t_{\mathrm{min}}+|t_{\mathrm{min}}|=0$), and there is an extra analog block of time $|\lambda_1 t_{\mathrm{min}}|$ not sandwiched by any SQR.  The evolution decomposes into
\begin{equation}
	U_{ZZ}(t_F)=e^{(i\sum_{\beta}g\mathsf{M}_{\alpha\beta}\tilde{t}_\alpha\sigma_z^j \sigma_z^k)}e^{(i|\lambda_1 t_{\mathrm{min}}| \sum_{\beta}g \sigma_z^j \sigma_z^k)},
\end{equation}
where the dependence of $j,k$ on $\beta$ has not been explicitly written and is the same as in Eq. (\ref{eq:H_beta}).

%

\section{\uppercase{Universality}}
\label{Appsec:Universality}
We briefly recall that a machine able to implement single qubit rotations and an entangling gate, e.g.,  a controlled-phase gate, has the ability to perform universal quantum computation efficiently. It can be easily proven that a two-qubit ZZ gate, $ZZ_{ij}(\Phi)=e^{-i \Phi \sigma_z^i \sigma_z^j}$, combined with single qubit rotations can be used to implement a controlled-phase gate $CZ_{ij}(\phi)=(Z_i(-\phi)\otimes Z_j(-\phi))ZZ_{ij}(\phi/2)$ (up to a general phase) where
\begin{equation}
Z(\phi)=\begin{pmatrix} 1 &0\\ 0 & e^{i\phi}\end{pmatrix},\,\, 
CZ(\phi)=\begin{pmatrix} 1 &0 &0 &0\\
0 &1 &0&0\\
0 & 0 &1&0\\
0&0&0& e^{-i2\phi}\end{pmatrix}.
\end{equation}

\section{\uppercase{Independent coupling parameters in XZ model}}
\label{Appsec:XZmodel_couplings}
The implementation of the XZ model with the protocol of Subsec. \ref{subsec:XZ_model} requires the solution of the following linear system
\begin{equation}
	\left[\alpha_j^{(\mu,s)}\alpha_k^{(\nu,s)}\right]\begin{pmatrix}
	g_{12}^{(1)}\\
	g_{12}^{(2)}\\
	g_{12}^{(3)}\\
	g_{12}^{(4)}\\
	g_{13}^{(1)}\\
	\vdots\\
	g_{N-1,N}^{(4)}
	\end{pmatrix}=\begin{pmatrix}
	g_{12}^{XX}\\
	g_{12}^{XZ}\\
	g_{12}^{ZX}\\
	g_{12}^{ZZ}\\
	g_{13}^{XX}\\
	\vdots\\
	g_{N-1,N}^{ZZ}
	\end{pmatrix},
\end{equation}
with the matrix  
\begin{eqnarray}
	\left[\alpha_j^{(\mu,s)}\alpha_k^{(\nu,s)}\right]=\bigoplus_{j<k}^N
	\begin{pmatrix} 
	S_{\theta_j}^1 S_{\theta_k}^1 & S_{\theta_j}^2S_{\theta_k}^2&S_{\theta_j}^3 S_{\theta_k}^3 & S_{\theta_j}^4 S_{\theta_2}^4\\
	S_{\theta_j}^1C_{\theta_k}^1 & S_{\theta_j}^2C_{\theta_k}^2&S_{\theta_j}^3 C_{\theta_k}^3 & S_{\theta_j}^4 C_{\theta_k}^4\\
	C_{\theta_j}^1S_{\theta_k}^1 & C_{\theta_j}^2S_{\theta_k}^2&C_{\theta_j}^3 S_{\theta_k}^3 & C_{\theta_j}^4 S_{\theta_k}^4\\
	C_{\theta_j}^1C_{\theta_k}^1 & C_{\theta_j}^2C_{\theta_k}^2&C_{\theta_j}^3 C_{\theta_k}^3 & C_{\theta_j}^4 C_{\theta_k}^4\\
	\end{pmatrix},\nonumber\\
\end{eqnarray}
where we have defined the parameters $S_{\theta_j}^s=\sin(\theta_j^{(s)})=\alpha_j^{(x,s)}$ and $C_{\theta_j}^s=\cos(\theta_j^{(s)})=\alpha_j^{(z,s)}$. This matrix is invertible for a dense set of phase values. That is, the sets of phases that make it singular has measure zero. From a practical perspective, we do not want eigenvalues close to zero, because after inversion we would have long  simulating times. One useful and well-behaved array of phases is $\theta_w^{(s)}=\frac{s\pi(w)}{2(w+1)}$, with distance between two nearest-neighbour qubit phases scaling polynomially $d(s,w)=|\theta_w^{(s)}-\theta_{w+1}^{(s)}|=\frac{s\pi}{2(w^2+3w+2)}$.

\section{\uppercase{$\mathbf{M}$-body Hamiltonians}}
\label{Appsec:Mbody}
\begin{figure}[h]
	\includegraphics[width=.9\linewidth]{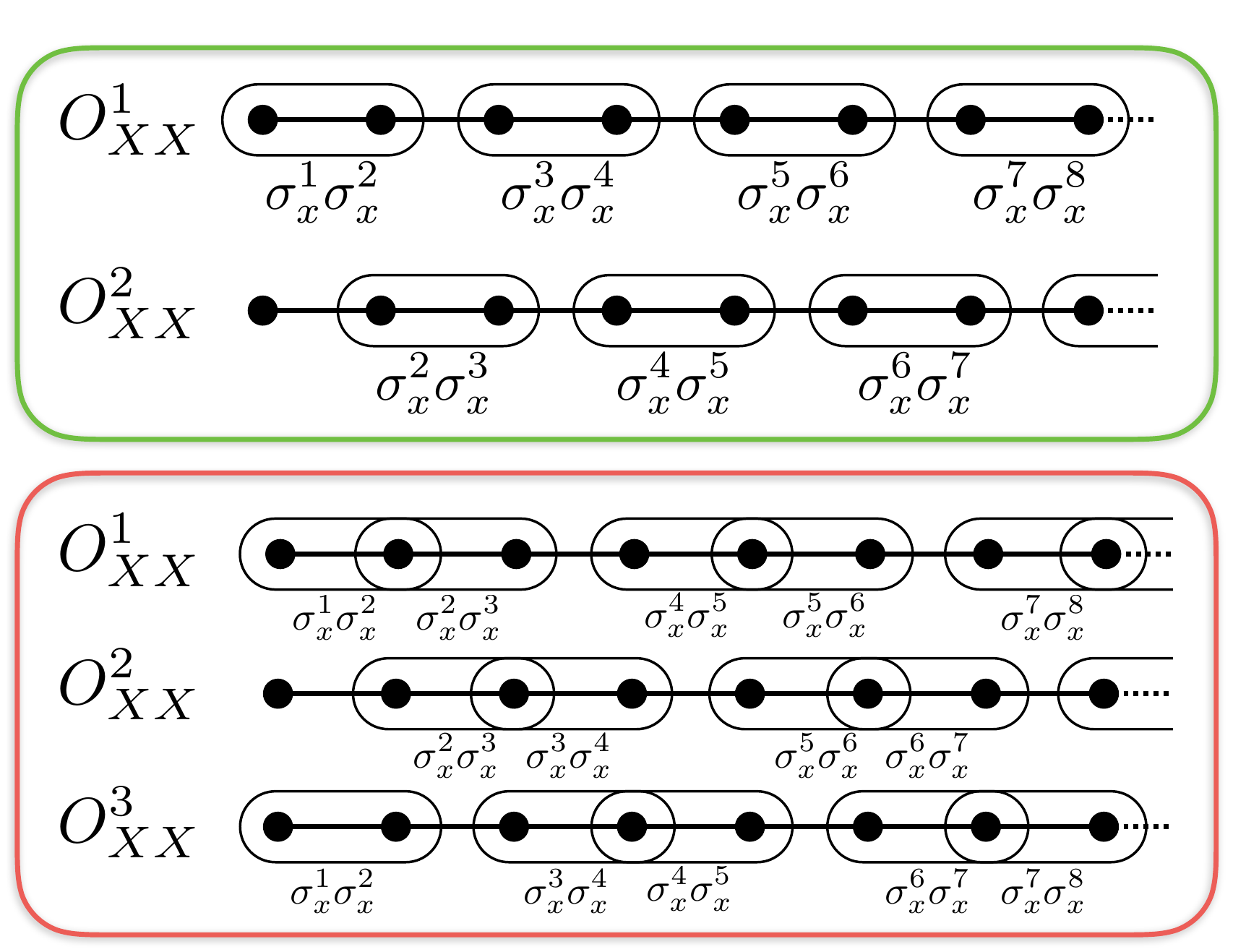}
	\caption{\label{fig:M_body_O_XX} \textbf{Sets of engineered interactions $O_{XX}^s$ for creating all-support interactions}. (Green) Two sets of generalized Hamiltonians required to create interactions in all supports for  $M\le4$. (Red) Three sets of generalized Hamiltonians required to create all interactions with support $M\le6$.}
\end{figure}

We extend here the steps explained in Subsec. \ref{Subsec:M_body} to simulate a NN Hamiltonian evolution with up to $4$-body interactions like 
\begin{eqnarray}
H_{4b}&=&\sum_{j,\chi\eta} g_{(2,j)}^{\chi\eta}\sigma_{\chi}^{j}\sigma_{\eta}^{j+1}+\sum_{j,\chi\eta\gamma} g_{(3,j)}^{\chi\eta\gamma}\sigma_{\chi}^{j}\sigma_{\eta}^{j+1}\sigma_{\gamma}^{j+2}\nonumber\\
&&+\sum_{j,\chi\eta\gamma\rho}g_{(4,j)}^{\chi\eta\gamma\rho}\sigma_{\chi}^{j}\sigma_{\eta}^{j+1}\sigma_{\gamma}^{j+2}\sigma_{\rho}^{j+3},
\end{eqnarray}
where $\{\chi,\eta,\gamma,\rho\}=\{x,y,z\}$ and $j=\{1,...,N\}$, starting with NN fixed coupling ZZ Ising models. To create terms with support in all interactions by a generalized M{\o}lmer-S{\o}rensen type of gate, we need to interleave inhomogeneous Ising Hamiltonians with two different and rotated XX-Ising evolutions as
\begin{eqnarray}
H_1=e^{-iO_{XX}^{1}}H_{ZZ}^1e^{iO_{XX}^{1}},\\
H_2=e^{-iO_{XX}^{2}}H_{ZZ}^2e^{iO_{XX}^{2}},
\end{eqnarray}
where $O_{XX}^{1}=\Phi_1\sigma_x^{1}\sigma_x^{2}+\Phi_3\sigma_x^{3}\sigma_x^{4}+\Phi_5\sigma_x^{5}\sigma_x^{6}+...$ and its translationally shifted $O_{XX}^{2}=\Phi_2\sigma_x^{2}\sigma_x^{3}+\Phi_4\sigma_x^{4}\sigma_x^{5}+\Phi_6\sigma_x^{6}\sigma_x^{7}+...$ are built from evolutions of ZZ models rotated with SQRs in all qubits $R=\otimes_j^N\left(\cos(\pi/4)\sigma_z^j+\sin(\pi/4)\sigma_x^j\right)$. For $M=4$, $O_{XX}^{1}$ and $O_{XX}^{2}$ contain interacting operators separated by the interaction length $L=M/2=2$, e.g.,  $O_{XX}^{1}$ has a term $\sigma_x^{1}\sigma_x^{2}$ but not $\sigma_x^{2}\sigma_x^{3}$, see Fig. \ref{fig:M_body_O_XX}. Had we wanted to simulate a five-/six- (seven-/eight-) body Hamiltonian, we would need a different decomposition with 3 (4) translationally invariant sets of blocks, see Fig. \ref{fig:M_body_O_XX}. The ZZ-Ising NN Hamiltonians $H_{ZZ}^{s}=\sum_j g_j^{s} \sigma_z^{j}\sigma_z^{j+1}$, with $s=\{1,2\}$. 

$H_1$ contains all two-body and three-body terms with different supports but not in four-body terms, i.e.,  for a chain of 8 qubits it looks like
\begin{eqnarray}
	H_1=&g_1^1\cos(2\theta_2)\sigma_z^1\sigma_z^2+g_1^1\sin(2\theta_2)\sigma_z^1\sigma_y^2\sigma_x^3\nonumber\\
	&+g_2^1\sigma_z^2\sigma_z^3 +g_3^1\cos(2\theta_2)\cos(2\theta_4)\sigma_z^3\sigma_z^4\nonumber\\
	&+g_3^1\sin(2\theta_2)\cos(2\theta_4)\sigma_x^2\sigma_y^3\sigma_z^4\nonumber\\
	&+g_3^1\cos(2\theta_2)\sin(2\theta_4)\sigma_z^2\sigma_y^3\sigma_x^4\nonumber\\
	&+g_3^1\sin(2\theta_2)\sin(2\theta_4)\sigma_x^2\sigma_y^3\sigma_y^4\sigma_x^5\nonumber\\
	&+g_4^1\sigma_z^4\sigma_z^5 +g_5^1\cos(2\theta_4)\cos(2\theta_6)\sigma_z^5\sigma_z^6\nonumber\\
	&+g_5^1\sin(2\theta_4)\cos(2\theta_6)\sigma_x^4\sigma_y^5\sigma_z^6\nonumber\\
	&+g_5^1\cos(2\theta_4)\sin(2\theta_6)\sigma_z^4\sigma_y^5\sigma_x^6\nonumber\\
	&+g_5^1\sin(2\theta_4)\sin(2\theta_6)\sigma_x^4\sigma_y^5\sigma_y^6\sigma_x^7\nonumber\\
	&+g_6^1\sigma_z^6\sigma_z^7 +g_7^1\cos(2\theta_6)\sigma_z^7\sigma_z^8\nonumber\\
	&+g_7^1\sin(2\theta_6)\sigma_x^6\sigma_y^7\sigma_z^8.\label{eq:M_body_H_1}	
\end{eqnarray}

On the other hand, $H_2$ contains (again) terms in all supports for two-body and three-body interactions and the complementary four-body terms 
\begin{eqnarray}
H_2=&g_1^2\sigma_z^1\sigma_z^2+g_2^2\cos(2\theta_1)\cos(2\theta_3)\sigma_z^2\sigma_z^3\nonumber\\
&+g_2^2\sin(2\theta_1)\cos(2\theta_3)\sigma_x^1\sigma_y^2\sigma_z^3\nonumber\\
&+g_2^2\cos(2\theta_1)\sin(2\theta_3)\sigma_z^2\sigma_y^3\sigma_x^4\nonumber\\
&+g_2^2\cos(2\theta_1)\cos(2\theta_3)\sigma_x^1\sigma_y^2\sigma_y^3\sigma_x^4\nonumber\\
&g_3^2\sigma_z^3\sigma_z^4+g_4^2\cos(2\theta_3)\cos(2\theta_5)\sigma_z^4\sigma_z^5\nonumber\\
&+g_4^2\sin(2\theta_3)\cos(2\theta_5)\sigma_x^3\sigma_y^4\sigma_z^5\nonumber\\
&+g_4^2\cos(2\theta_3)\sin(2\theta_5)\sigma_z^4\sigma_y^5\sigma_x^6\nonumber\\
&+g_4^2\cos(2\theta_3)\cos(2\theta_5)\sigma_x^3\sigma_y^4\sigma_y^5\sigma_x^6\nonumber\\
&g_5^2\sigma_z^5\sigma_z^6+g_6^2\cos(2\theta_5)\cos(2\theta_7)\sigma_z^6\sigma_z^7\nonumber\\
&+g_6^2\sin(2\theta_5)\cos(2\theta_7)\sigma_x^5\sigma_y^6\sigma_z^7\nonumber\\
&+g_6^2\cos(2\theta_5)\sin(2\theta_7)\sigma_z^6\sigma_y^7\sigma_x^8\nonumber\\
&+g_6^2\sin(2\theta_5)\sin(2\theta_7)\sigma_x^5\sigma_y^6\sigma_y^7\sigma_x^8\nonumber\\
&+g_7^2\sigma_y^7\sigma_z^8.\label{eq:M_body_H_2}
\end{eqnarray}

The constant coefficients of operators in (\ref{eq:M_body_H_1}) and (\ref{eq:M_body_H_2}) are entangled in groups of maximum size 4. For the simulation of the four-body generalized Hamiltonian, it is thus enough to sum 4 of each of the blocks $H_0=\sum_{k=1}^4 H_1^{(k)}+H_2^{(k)}$ to decouple the parameters of at least one term operating in each support. We have again a dense set of phases such that randomly chosen ones would make the system of equations invertible. A particular choice of sets that would work are $\theta_1^{(k)}=\theta_{1+4n}^{(k)}=\theta_2^{(k)}=\theta_{2+4n}^{(k)}=(2\pi k/3)$ and $\theta_3^{(k)}=\theta_{3+4n}^{(k)}=\theta_4^{(k)}=\theta_{4+4n}^{(k)}=(2\pi k/5)$ with $n=\{1,\dots,[N/4]\}$ and $k=\{1,\dots4\}$. 

Finally, we use more local rotations to generate the arbitrary $M\le4$ Hamiltonian. In particular, we need to concatenate $3^{M}=81$ $H_0$ blocks, maximum number of independent parameters in an $M$-body interaction,interleaved by generalized SQRs $R^{(l)}=\otimes_j^N r_j^{(l)}\sigma_x^{j}+s_j^{(l)}\sigma_y^{j}+t_j^{(l)}\sigma_z^{j}$
\begin{equation}
H_{4b}=\sum_l R^{(l)} H_0^{(l)} R^{(l)},
\end{equation}
where $(r_j^{(l)},s_j^{(l)},t_j^{(l)})$ are unit-sphere cartesian decompositions that fulfill the constraint $|r_j^{(l)}|^2+|s_j^{(l)}|^2+|t_j^{(l)}|^2=1$. As it is common when simulating non-commuting Hamiltonians, we need to repeat the whole process for each Trotter step $e^{-iH_{4b}t}\approx (e^{-iH_{4b}t/n_T})^{n_T}$ to approximate the evolution.

\end{document}